\documentclass[paper,11pt]{article}
\pdfoutput=1

\usepackage{mathrsfs}
\usepackage{graphicx}
\usepackage{epstopdf}
\usepackage{color}
\usepackage[centertags]{amsmath}
\usepackage{amsfonts}
\usepackage{amssymb} 
\usepackage{amsthm}
\usepackage{slashed}
\usepackage{framed}
\usepackage{enumerate,enumitem}
\usepackage{amsbsy}
\usepackage{jheppub}
\usepackage{subfig}
\usepackage{hyperref}
\usepackage{braket}
\usepackage{bbm}
\usepackage{comment}

\def\beq{\begin{equation}}
\def\eeq{\end{equation}}
\def\beqa{\begin{eqnarray}}
\def\eeqa{\end{eqnarray}}
\newcommand{\eqa}{\begin{eqnarray}}
\newcommand{\ena}{\end{eqnarray}}

\newcommand{\ta}{\tau}
\newcommand{\uu}{\phi}
\newcommand{\cA}{\mathcal{A}}
\newcommand{\x}{\chi}
\newcommand{\MBI}{\text{\tiny \bf MBI}}
\newcommand{\Max}{\text{\tiny \bf M}}
\newcommand{\YM}{\text{\tiny \bf YM}}
\newcommand{\cR}{\mathcal{R}}
\newcommand{\cM}{\mathcal{M}}
\newcommand{\bC}{{\bf C}}

\title{\boldmath Generalised Born-Infeld   models, Lax operators  and the $\textsc{T} \bar{\textsc{T}}$ perturbation }

\author[1]{Riccardo Conti,}
\author[1]{Leonardo Iannella,}
\author[2]{Stefano Negro,}
\author[1]{Roberto Tateo}

\affiliation[1]{Dipartimento\ di Fisica and INFN, Universit\`a di Torino, Via P.\ Giuria 1, 10125 Torino, Italy.}
\affiliation[2]{ C.N. Yang Institute for Theoretical Physics, New York Stony Brook, NY 11794-3840. U.S.A.}

\emailAdd{riccardo.conti@to.infn.it}
\emailAdd{leonardiannell@gmail.com}
\emailAdd{steff.negro@gmail.com}
\emailAdd{tateo@to.infn.it}

\abstract{Surprising links between the deformation of 2D quantum field theories induced by the composite $\textsc{T} \bar{\textsc{T}}$ operator, effective string models and the $AdS/$CFT correspondence, have recently emerged. 
The purpose of this article is to discuss various classical aspects related to the deformation of 2D interacting field theories. 
Special attention is given to the  sin(h)-Gordon model,  for which we were able to construct the $\textsc{T} \bar{\textsc{T}}$-deformed Lax pair. 
We consider the Lax pair formulation to be  the first essential step  toward a more satisfactory geometrical interpretation of this deformation within the integrable model framework. 

Furthermore, it is  shown that the 4D Maxwell-Born-Infeld theory, possibly with the addition of a mass term or a derivative-independent potential, corresponds to a natural extension of the 2D examples. Finally, we  briefly comment on 2D Yang-Mills theory and propose a modification of the heat kernel, for a generic surface with genus $p$ and $n$ boundaries,  which fully accounts for the $\textsc{T} \bar{\textsc{T}}$ contribution. 
}

\begin{document} 
\maketitle
\flushbottom

\section{Introduction }
\label{sec:intro}
Effective  Field Theories (EFTs), are characterized by  the presence of  irrelevant fields in the Lagrangian  which  usually  make quantization and the  physical interpretation of the high-energy  regime very problematic.  In two spacetime dimensions,  the  study  of  EFTs  is experiencing a period of renewed interest thanks to the discovery  of  surprising integrable-like   properties of the  $\textsc{T} \bar{\textsc{T}}$ composite operator,  rigorously defined by Zamolodchikov  \cite{Zamolodchikov:2004ce} as the determinant of the stress-energy tensor. 

While the main source of inspiration of \cite{Zamolodchikov:2004ce} were the non-perturbative factorization properties detected, within the Form-Factor approach, in \cite{Fateev:1997yg}, the $\textsc{T} \bar{\textsc{T}}$ perturbative contributions  to the finite-size spectrum  first emerged  from the   study of the  RG flow  connecting the  Tricritical Ising  (TIM)  to the Ising model (IM) \cite{Zamolodchikov:1991vx}. 
The analysis of  \cite{Zamolodchikov:1991vx}, was based on  a combination of powerful techniques such as  conformal perturbation theory, exact scattering theory and  the  Thermodynamic Bethe Ansatz (TBA).  

The scattering  among  right and left mover massless excitations  along the  TIM $\rightarrow $ IM  critical line is described  by a pure CDD \cite{Castillejo:1955ed} factor which, therefore, should contain information on irrelevant fields. This observation triggered   early  studies on  TBA models  with modified CDD  kernels  and lead to the conclusion that, in many cases, they were affected by  short-distance instabilities \cite{Alzam, Mussardo:1999aj} (see the related  discussion in  Section  9 of \cite{Smirnov:2016lqw}).  The fact that  seemingly  consistent exact  S-matrix models\footnote{For example, the wide family of scattering  models proposed in the final discussion Section of  \cite{Hollowood:1993ac}.}  may display ultraviolet pathological behavior was first detected in \cite{Ravanini:1992fi}.
The interest towards this research topic remained very limited for many years  until an important step forward was made in \cite{dubovsky2012solving, dubovsky2012effective}: a link between the TBA equations for free massless bosons,  modified by a specific CDD factor, and the spectrum of  effective bosonic closed strings  was discovered. 
The generalization to open strings, to other conformal field theories  and the observation  that the effective action describing the confining flux tube of a generic gauge theory was described, at least at leading order,  by a $\textsc{T} \bar{\textsc{T}}$ perturbation was made in \cite{Caselle:2013dra}. 
The connection between these observations and the paper \cite{Zamolodchikov:2004ce}   was further clarified  in \cite{Smirnov:2016lqw, Cavaglia:2016oda} where, among many other results, an inviscid Burgers equation  for the spectrum  was identified, and the corresponding   equation for the action \cite{Smirnov:2016lqw} lead to the reconstruction of the whole  bosonic  Born-Infeld (BI)  Lagrangian in 2D \cite{Cavaglia:2016oda}. 

Triggered by these works, remarkable connections  have emerged  with the $AdS/$CFT duality \cite{McGough:2016lol,Turiaci:2017zwd, Giveon:2017nie, Giveon:2017myj,Asrat:2017tzd,Giribet:2017imm,Kraus:2018xrn, Cottrell:2018skz,  Baggio:2018gct,Babaro:2018cmq} and flat  space  Jackiw-Teitelboim  (JT)  gravity \cite{ Dubovsky:2017cnj,  Dubovsky:2018bmo}, together with generalizations to non Lorentz-invariant perturbations \cite{Bzowski:2018pcy, Guica:2017lia,Chakraborty:2018vja,Apolo:2018qpq}.

The study of partition functions of $\textsc{T} \bar{\textsc{T}}$-deformed models was started in  \cite{Cavaglia:2016oda} and further developed in \cite{Cardy:2018sdv,  Dubovsky:2018bmo, Datta:2018thy}.\footnote{See also \cite{Luscher:2004ib,Billo:2006zg} for earlier results on partition functions for the bosonic Born-Infeld models,  in the context of effective flux-tube theories.}
Interesting  results  on entanglement were recently obtained  in \cite{Chakraborty:2018kpr, Donnelly:2018bef}. 
Finally, a link with stochastic processes was established and  generalizations to higher spacetime dimensions  proposed  in \cite{Cardy:2018sdv} (see also in \cite{Bonelli:2018kik, Taylor:2018xcy}). 

The purpose of this article is to further investigate the properties of $\textsc{T} \bar{\textsc{T}}$-deformed field theories.  Firstly, we shall review some of the results reported in \cite{Cavaglia:2016oda}, concerning classical bosonic Lagrangians with interacting  potentials.  
We will prove that the fairly complicated expression for the perturbed Lagrangian, given  in \cite{Cavaglia:2016oda},  can be recast  into  a much  simpler Born-Infeld type form. 
We shall also  comment on the similarity between the inclusion of the potential term and a transformation property for the spectrum first spotted in  \cite{Smirnov:2016lqw}, as the  coefficient of the bulk contribution of the unperturbed energy is modified. The latter  results were  anticipated in  \cite{RTTalkIGST2017} 
and  are partially connected, with  some minor overlap,  to the  papers \cite{Baggio:2018gct,Bonelli:2018kik}. The $\textsc{T} \bar{\textsc{T}}$-deformed  sine-Gordon model is also discussed in detail and the  corresponding Lax operators are constructed.

Furthermore, motivated  by the observations made many years ago in \cite{1966JETP, Barbashov:1967zzz} which  link plane wave scatterings in the 4D Maxwell-Born-Infeld (MBI) theory to a 2D bosonic Born-Infeld model,  we shall show that the  MBI Lagrangian satisfies a simple generalization of the equations described in \cite{Smirnov:2016lqw, Cavaglia:2016oda}, similar  but different from the higher dimensional  proposals of \cite{Cardy:2018sdv,Bonelli:2018kik,Taylor:2018xcy}. The introduction  of a mass term  or a derivative independent potential in the original field theory affects the  $\textsc{T} \bar{\textsc{T}}$-deformed Lagrangian as in the  2D examples.

Finally, we  will briefly discuss the exactly solvable example of 2D Yang-Mills and conjecture a simple modification that includes the $\textsc{T} \bar{\textsc{T}}$ contribution in the partition functions, and more generally in the  heat kernel for a generic surface with genus $p$ and $n$ boundaries.

\section{Deformed interacting bosonic Lagrangians from the Burgers equation}
\label{sec:DefLag}
In \cite{Smirnov:2016lqw, Cavaglia:2016oda}  it was proven that the energy levels $E_n(R,\ta)$ associated to the stationary states $\ket{n}$ with spatial momenta $P_n(R) = \frac{2\pi k_n}{R}$,  ($k_n\in\mathbb{Z}$), satisfy the following inhomogeneous Burgers equation
\beq
\label{eq:Burgers}
\partial_{\ta} E_n(R,\ta) = \frac{1}{2} \partial_R\left( E_n^2(R,\ta) - P_n^2(R) \right)= -\frac{R}{\pi^2} \braket{n|\textsc{T} \bar{\textsc{T}}|n}_R\;,
\eeq
where the composite operator $\textsc{T} \bar{\textsc{T}}$ is defined up to total derivative terms as
\beq
\textsc{T} \bar{\textsc{T}}(z,\bar{z}) := \lim_{(z',\bar{z}')\rightarrow(z,\bar{z})}  T(z,\bar{z})\bar{T}(z',\bar{z}')-\Theta(z,\bar{z})\Theta(z',\bar{z}') \;,
\eeq
and the complex components $T$, $\bar{T}$ and $\Theta$ of the stress-energy tensor are related to the Euclidean components $T_{11}$, $T_{22}$ and $T_{12}$   by the following relations:
\beq
(x_1,x_2) = (x,t) \;,\; (z,\bar{z})=  (x_1+\mathbbm{i}\,x_2 \;,\;  x_1-\mathbbm{i}\,x_2) \;,
\eeq
\beq
T_{11} = -\frac{1}{2\pi}(\bar{T}+T-2\Theta) \;,\; T_{22} = \frac{1}{2\pi}(\bar{T}+T+2\Theta) \;,\; T_{12} = T_{21}  =\frac{\mathbbm i}{2\pi}(\bar{T}-T) \;.
\eeq
At finite volume $R$, the expectation values of the Euclidean components of the stress-energy tensor are related to $E_n$ and $P_n$  through \cite{ZamolodchikovTBA}:
\beq
\label{eq:EvsT}
E_n(R,\ta) = -R\bra{n}T_{22}\ket{n} \;,\; \partial_R E_n(R,\ta) = -\bra{n}T_{11}\ket{n} \;,\; P_n(R) = -\mathbbm i R\bra{n}T_{12}\ket{n} \;.
\eeq
Since (\ref{eq:Burgers}) holds for any $n$, in the following we will drop the subscript $n$: $E_n(R,\ta) = E(R,\ta)$ and $P_n(R) = P(R) = \frac{2\pi k}{R} \;,\; (k\in\mathbb{Z})$. As a side remark, notice that from \cite{Cavaglia:2016oda} it follows
\beq
\label{eq:Lorentz}
 \left(\begin{array}{c}
E(R,\ta) \\
P(R)
\end{array}\right) =
\left(\begin{array}{cc}
\cosh{(\theta_0)} & -\sinh{(\theta_0)} \\
-\sinh{(\theta_0)} & \cosh{(\theta_0)} 
\end{array}\right) 
\left(\begin{array}{c}
E(\cR_0,0) \\
P(\cR_0)
\end{array}\right) \;,
\eeq
with
\beq
\sinh{\theta_0} = \frac{\ta\,P(R)}{\cR_0} = \frac{\ta\,P(\cR_0)}{R} \;,\; \cosh{\theta_0} = \frac{R + \ta\,E(R,\ta)}{\cR_0} = \frac{\cR_0 + \ta\,E(\cR_0,0)}{R} \;,
\eeq
and
\beq
\cR_0^2 = \left(R+\ta\,E(R,\ta) \right)^2 - \ta^2 P^2(R) \;,\; R^2 = \left(\cR_0+\ta\,E(\cR_0,0) \right)^2 - \ta^2 P^2(\cR_0) \;.
\eeq
Therefore the solution to (\ref{eq:Burgers}) can be written in implicit form as
\beq
E^2(R,\ta) - P^2(R) = E^2(\cR_0,0) - P^2(\cR_0,0) \;.
\eeq
It would be interesting to check if there exists an extension to higher spacetime dimensions of the Lorentz-type map (\ref{eq:Lorentz}) corresponding to the generalizations of the $\textsc{T}\bar{\textsc{T}}$ deformation proposed in \cite{Cardy:2018sdv,Bonelli:2018kik,Taylor:2018xcy} and/or to the quantum version of the Maxwell-Born-Infeld model discussed in Section \ref{sec:MBI}. \\
If the boundary conditions at $\ta=0$ are the energy levels of a CFT, {\it i.e.} of the form:
\beq
\label{eq:boundenergy1}
E(R,0) = \frac{A}{R} \;,
\eeq
the general solution to (\ref{eq:Burgers}) is
\beq
\label{eq:energy1}
E(R,\ta) = \frac{R}{2 \ta} \left( -1 + \sqrt{1 + \frac{4 \ta}{R^2} A + \frac{4\ta^2}{R^2} P^2(R)} \right) = \frac{R}{2 \ta} \left( -1 + \sqrt{1 + \frac{4 \ta}{R^2}A + \frac{4\ta^2}{R^4}(2\pi k)^2}\right) \;.
\eeq
The consequence, on the latter expression, of an additional bulk term  in the unperturbed energy (\ref{eq:boundenergy1}),
\beq
E(R,0) = \frac{A}{R} + F_0 R \;,
\label{eq:initialE1}
\eeq
was considered in \cite{Smirnov:2016lqw}. Imposing the  initial condition (\ref{eq:initialE1}), the solution to (\ref{eq:Burgers}) becomes: 
\beq
E(R,\ta) = \frac{F_0 R}{1-\ta\,F_0} + \frac{R}{2\tilde{\ta}} \left( -1 + \sqrt{1 + \frac{4\tilde{\ta}}{R^2}A + \frac{4\tilde{\ta}^2}{R^2}P^2(R)} \right) \;,
\label{eq:EBulk}
\eeq
with $\tilde{\ta} = \ta(1-\ta F_0)$,  that is a reparametrization  
$\Delta E_n(R,\ta) \rightarrow \Delta E_n(R,\tilde{\ta})$
of the perturbing parameter $\ta$ in the energy differences $\Delta E_n(R,\ta)=E_n(R,\ta)-E_0(R,\ta)$.\\
Furthermore,  it was argued in \cite{Smirnov:2016lqw} that (\ref{eq:Burgers}) is equivalent, up to total derivative terms, to the following fundamental equation for the Lagrangian :
\beq
\partial_{\ta} \mathcal{L}(\ta) =\text{det}[T_{\mu\nu}(\ta)] \;,\; \textsc{T} \bar{\textsc{T}}(\ta) =- \pi^2 \text{det}[T_{\mu\nu}(\ta)] \;,
\label{eq:LagTT0}
\eeq
with $\mu,\nu \in \lbrace 1,2\rbrace$ and Euclidean coordinates  $(x_1,x_2)$.
By solving perturbatively (\ref{eq:LagTT0}) with initial condition 
\beq
\label{eq:Nbos}
\mathcal{L}(\vec{\phi},  0) = \partial\vec{\phi}\cdot\bar{\partial}\vec{\phi} \;,\;\;\; \vec{\phi}=\left(\phi_1(z,\bar z),\dots  ,\phi_N(z,\bar z) \right) \;,
\eeq
it was proved in  \cite{Cavaglia:2016oda} that the deformed Lagrangian $\mathcal{L}(\vec{\phi}, \ta)$ coincides with the bosonic Born-Infeld model or, equivalently, the Nambu-Goto Lagrangian in the static gauge:  
\beq
\label{eq:NGL}
\mathcal{L}(\vec{\phi}, \ta) = \frac{1}{2 \ta} \left( -1+\sqrt{1+4 \ta \mathcal{L}(\vec{\phi},0)-4\ta^2\mathcal{B}}  \right)=
\frac{1}{2\ta}\left( -\sqrt{\det[\eta_{\mu\nu}]} + \sqrt{\det\left[ \eta_{\mu\nu} + \ta\, h_{\mu\nu} \right]} \right) \;,
\eeq
with $h_{\mu\nu}=\partial_{\mu}\vec{\phi}\cdot\partial_{\nu}\vec{\phi}$ and
\beq
\mathcal{B} = | \partial \vec{\phi} \times \bar{\partial} \vec{\phi} |^2= -\frac{1}{4}\det\left[ h_{\mu\nu} \right] \;.
\eeq
Here, we would like to extend the result (\ref{eq:NGL}) to generic interacting bosonic Lagrangians of the form:
\beq
\label{eq:LagV}
\mathcal{L}^V(\vec{\phi},0) = \partial\vec{\phi}\cdot\bar{\partial}\vec{\phi}+V(\vec{\phi}) \;,
\eeq
where $V(\vec{\phi})$ is a generic derivative-independent potential. Instead of solving  (\ref{eq:LagTT0}) using a perturbative brute-force approach, as in  \cite{Cavaglia:2016oda}, we proceed  by postulating that the evident  similarity between equations (\ref{eq:energy1}) and  (\ref{eq:NGL}), may be extended also to the $\textsc{T} \bar{\textsc{T}}$-deformation of (\ref{eq:LagV}).   
Concretely,  by comparing (\ref{eq:NGL}) with (\ref{eq:energy1}), it is easy to check that the following rescaled Lagrangian
\beq
\mathcal{L}_\x (\vec{\phi},\ta) = \frac{1}{\x}\mathcal{L}\left(\vec{\phi}, \frac{\ta}{\x^2} \right) \;,
\eeq
also satisfies a Burgers equation 
\beq
\label{eq:BurgersL}
\partial_{\ta} \mathcal{L}_\x (\vec{\phi}, \ta) = \mathcal{L}_\x (\vec{\phi},\ta)\,\partial_\x \mathcal{L}_\x (\vec{\phi},\ta) -\frac{\mathcal{B}}{\x^3} \;,
\eeq
with initial condition $\mathcal{L}_\x(\vec{\phi}, 0) =  \frac{1}{\x}\,\partial\vec{\phi}\cdot\bar{\partial}\vec{\phi}$.
Notice that the introduction of the auxiliary adimensional scaling parameter $\x$ allows us to establish a link between (\ref{eq:LagTT0}), {\it i.e.}
\beq
\partial_{\ta} \mathcal{L}_\x(\vec{\phi},\ta) = - \frac{1}{\pi^2}\frac{1}{\x}\textsc{T} \bar{\textsc{T}}_\x(\ta) \;,\; \textsc{T} \bar{\textsc{T}}_\x(\ta) = - \pi^2 \text{det}[T^{\mu\nu}_\x(\ta)] \;,
\label{eq:LagTT}
\eeq
and the Burgers equation  (\ref{eq:BurgersL}) for $\mathcal{L}_\x (\vec{\phi},\ta)$.
Motivated by this simple observation, we solve now (\ref{eq:BurgersL}) with  $\ta=0$  initial condition
\beq
\label{eq:boundL2}
\mathcal{L}_\x^V(\vec{\phi},0) = \mathcal{L}_\x(\vec{\phi}, 0) + \x\,V(\vec{\phi}) \;,
\eeq
the result is
\beq
\mathcal{L}_\x^V(\vec{\phi},\ta) =\frac{\x\,V(\vec{\phi})}{1-\ta\,V(\vec{\phi})} + \frac{\x}{2\bar{\ta}} \left( -1+\sqrt{1+\frac{4\bar{\ta}}{\x^2}\mathcal{L}(\vec{\phi},  0) -\frac{4\bar{\ta}^2}{\x^4}\,\mathcal{B} } \right) \;,
\label{eq:LV}
\eeq
with $\bar{\ta} = \ta(1-\ta V(\vec{\phi}))$. It is now  straightforward  to check that $\mathcal{L}_\x^V(\vec{\phi},\ta)$  still fulfills the fundamental equation (\ref{eq:LagTT}).\\
In the  $N=1$ case,  we first obtained the compact form (\ref{eq:LV}) performing a resummation of the more complicated, but equivalent, expression given in \cite{Cavaglia:2016oda} and subsequently we developed the more direct approach, which again maps (\ref{eq:LagTT}) to a Burgers-type equation. The latter technique was independently proposed in \cite{Bonelli:2018kik} and applied to different classes of systems and also to models in higher spacetime dimensions.
We address the interested reader to \cite{Bonelli:2018kik} for a detailed description of this alternative method.
The result (\ref{eq:LV}) is in perfect agreement with \cite{Dubovsky:2013ira}, where the first two perturbative contributions  of the  deformed free massive boson  action were determined using diagrammatic techniques.

It is also instructive to derive the classical Hamiltonian density $\mathcal{H}^V(\vec{\phi},\vec{\pi},\ta)$ associated to the Lagrangian density $\mathcal{L}^V(\vec{\phi},\ta)=\mathcal{L}_{\x=1}^V(\vec{\phi},\ta)$ and compare it with the expression of the quantized energy spectrum (\ref{eq:EBulk}). Using the shorthand notation $\vec{\phi}'=\partial_{1}\vec{\phi}$ and $\dot{\vec{\phi}}=\partial_{2}\vec{\phi}$ for the derivatives w.r.t. the Euclidean space and time respectively, the conjugated momentum is
\beq
\vec{\pi} = \frac{\partial\mathcal{L}^V(\vec{\phi},\ta)}{\partial\dot{\vec{\phi}}} \;,
\eeq
and the Hamiltonian density is a straightforward generalization of the single boson case reported in \cite{Kraus:2018xrn}
\beq
\label{eq:HamNG}
\mathcal{H}^V(\vec{\phi},\vec{\pi},\ta) = \frac{V(\vec{\phi})}{1-\ta\,V(\vec{\phi})} + \frac{1}{2\bar{\ta}}\left( -1 + \sqrt{1 + 4\bar{\ta}\,\mathcal{H}(\vec{\phi},\vec{\pi},0) + 4\bar{\ta}^2\,\mathcal{P}^2(\vec{\phi},\vec{\pi})} \right) \;,
\eeq
where $\mathcal{H}(\vec{\phi},\vec{\pi},0) = \frac{1}{4} |\vec{\phi}'|^2 - |\vec{\pi}|^2 = - T_{22}(0)$ is formally the Hamiltonian density of the free undeformed theory, while $\mathcal{P}(\vec{\phi},\vec{\pi}) = -{\mathbbm i}\,\vec{\pi}\cdot\vec{\phi}' = -{\mathbbm i}\,T_{12}(\ta)$ is the conserved momentum density of the deformed theory, following the convention (\ref{eq:EvsT}).\\
Notice that expression (\ref{eq:HamNG}) has the same formal structure of (\ref{eq:EBulk}). It is then easy to show that, introducing the auxiliary variable $\x$ in $\mathcal{H}^V(\vec{\phi},\vec{\pi},\ta)$ exactly in the same way as in $\mathcal{L}^V(\vec{\phi},\ta)$, the Hamiltonian density fulfills an inhomogeneous Burgers equation analogous to (\ref{eq:Burgers}) with the replacements
\beq
R\rightarrow \x \;,\; P^2\rightarrow \mathcal{P}^2 \;.
\eeq
Finally let us make some concluding remarks concerning the structure of the energy spectrum (\ref{eq:EBulk}). Looking at expression (\ref{eq:EBulk}), we notice the appearance of new special points in the parameter $\ta$, beside the square-root singularity already discussed in \cite{dubovsky2012solving, dubovsky2012effective, Caselle:2013dra, Smirnov:2016lqw, Cavaglia:2016oda}.

\begin{itemize}
\item The deformed bulk term $F(\ta) = \frac{F_0 R}{1-\ta\,F_0}$ in (\ref{eq:EBulk}) diverges at $\ta_{\text{LP}} = \frac{1}{F_0}$ which represents a Landau-type pole  singularity.
\item There exists a unique value $\ta_0 = \frac{1}{2F_0}$ such that the energy spectrum reduces exactly to a pure square-root form, without any additional term
\beq
E(R,\ta_0) = \frac{R}{2\tilde{\ta_0}} \sqrt{1 + \frac{4\tilde{\ta_0}}{R^2}A + \frac{4\tilde{\ta_0}^2}{R^2}P^2(R)} \;,\; \tilde{\ta}_0 = \ta_0(1-\ta_0 F_0) \;.
\eeq
\end{itemize}
As noticed in \cite{Caselle:2013dra}, in this case the finite-size expectation value of the $\textsc{T} \bar{\textsc{T}}$ becomes size and state independent:
\beq
\braket{\textsc{T} \bar{\textsc{T}}(\tau_0)}_R= -\frac{\pi^2}{2 R} \partial_{R} \left( E^2(R,\ta_0)-P^2(R) \right)= -\left(\frac{\pi}{2 \tilde{\ta}_0} \right)^2 \;.
\eeq
Here we would like to make the additional remark that, with the choice of a  constant potential $V(\vec{\phi})= F_0$ in (\ref{eq:LV}), the $\textsc{T} \bar{\textsc{T}}$ composite field  becomes  $\vec{\phi}$-independent at  $\ta=\ta_0$ : 
\beq
\textsc{T} \bar{\textsc{T}}(\ta_0) = -\left(\frac{\pi}{2 \tilde{\ta}_0} \right)^2 \;.
\eeq

\section{The $\textsc{T} \bar{\textsc{T}}$-deformed  sine-Gordon model}
\label{sec:sine}
Out of all  possible bosonic  theories corresponding to the Lagrangian density (\ref{eq:LV}), in this Section we will focus on the $\textsc{T} \bar{\textsc{T}}$-deformed classical sine-Gordon model, which corresponds to the case of a single boson field $\phi$ interacting with a sine potential. We will first derive the exact expression of the single kink solution at any value of the perturbing parameter $\ta$ and discuss the effect of the deformation, as $\ta$ is varied. The main  result of this Section  is the proof that the $\textsc{T} \bar{\textsc{T}}$ deformation preserves the classical integrability of the sine-Gordon model, we will arrive to this conclusion by explicitly constructing  the Lax pair of the deformed theory.
\subsection{Simple kink-like solutions}
\label{sec:kink}
\begin{figure}
 \centering
  \subfloat[]{\includegraphics[scale=0.23]{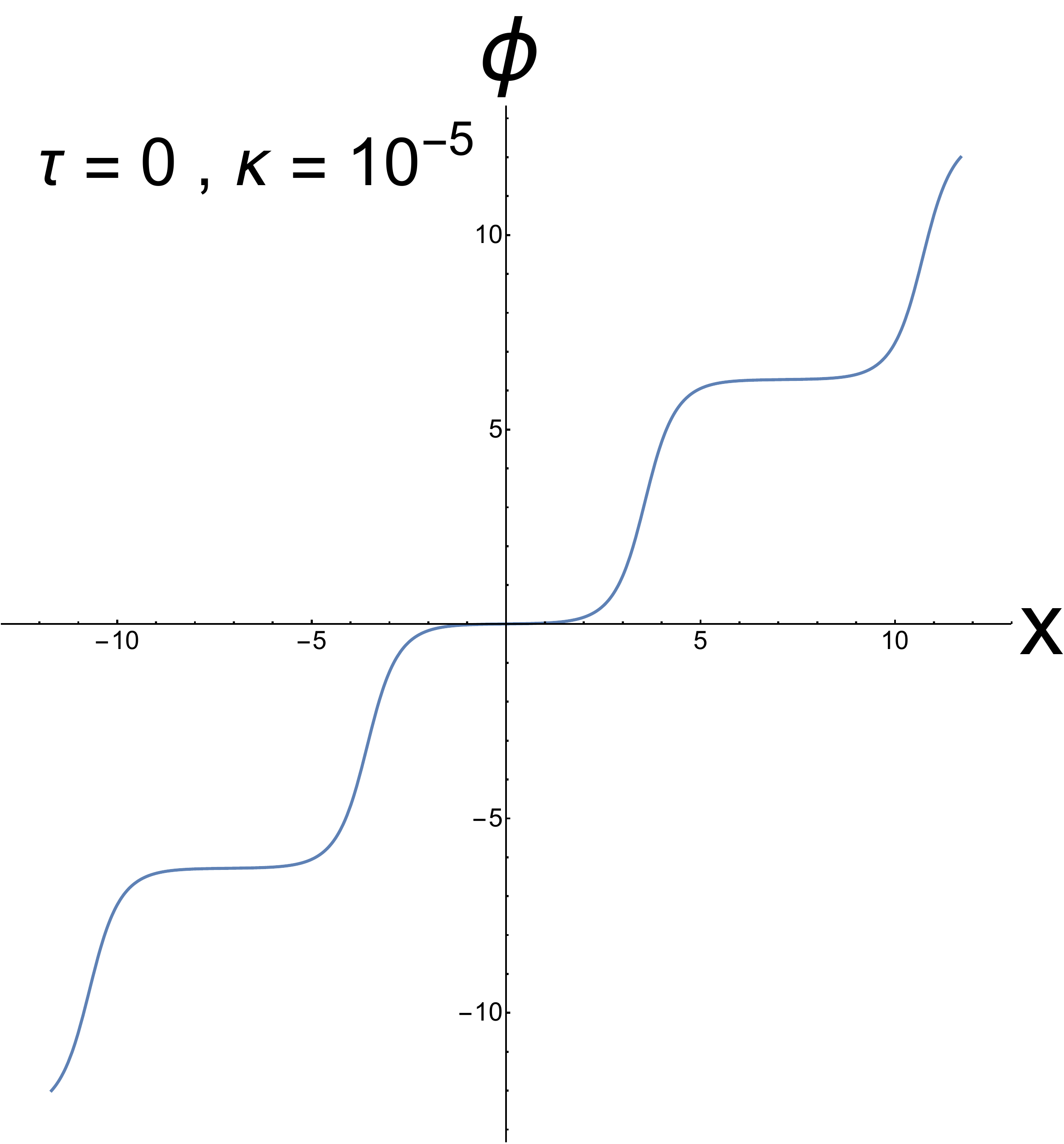}\label{fig:ke-5}}
  \hspace{1cm}
  \subfloat[]{\includegraphics[scale=0.23]{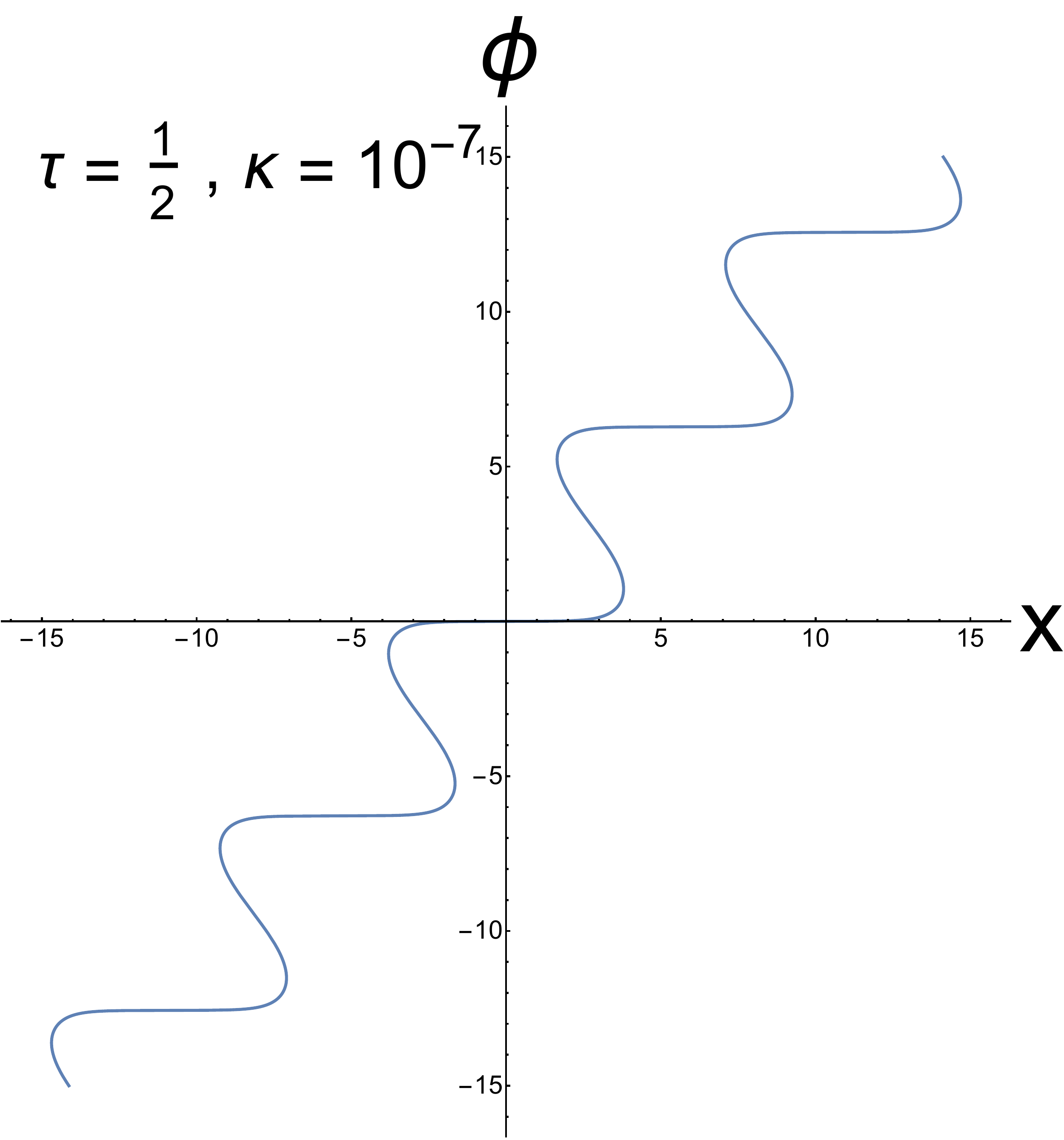}\label{fig:ke-7}}
  \vspace{0.5cm}
   \caption{The general solution (\ref{eq:elsolu}) for the undeformed (a) and the deformed (b) theory, for small values of $\kappa$.}
  \label{fig:Stair}
\end{figure}

Consider the sine-Gordon Lagrangian in Minkowski coordinates $(x,t)$ with signature $\eta_{\mu\nu} = \text{diag}(+1,-1)$ defined as
\beq
\mathcal{L}_{\text{SG}}(\phi) = \frac{1}{4}(\phi_x^2-\phi_t^2) + V\left(\uu\right) \;,\; V\left(\uu\right)= 4\sin^2(\uu/2) \;,
\label{eq:sGpotential}
\eeq
and the $\textsc{T} \bar{\textsc{T}}$-deformed sine-Gordon Lagrangian
\beq
\label{eq:sGDef}
\mathcal{L}_{\text{SG}}(\phi,\ta) = \frac{V}{1-\ta V}+\frac{1}{2\ta\left(1-\ta V\right)}\left( -1 + \sqrt{1 + \ta\,(1-\ta\,V)(\phi_x^2-\phi_t^2)} \right) \;,
\eeq
where the shorthand notation $\phi_\mu = \partial_\mu\phi$ for spacetime derivatives will be used hereafter.\\
The equations of motion (EoMs) associated to (\ref{eq:sGDef}) can be compactly written as
\beqa
\label{eq:EoMs}
\left(1-\tau V\right)^{2}\left(\uu_{xx} -\uu_{tt}\right)&-&\tau\left(1-\tau V\right)^{3}\left(\uu_{xx}\uu_{t}^{2}-2\uu_{xt} \uu_{x}\uu_{t}+\uu_{tt}\uu_{x}^{2}\right)  \notag \\
&=&\frac{1}{2}\tau V'\left(1-\tau V\right)\left(3+2S\right)\left(\uu_{x}^{2}-\uu_{t}^{2}\right)+\left(1+S\right)V'\;,
\eeqa
where we have set
\beq
S=\sqrt{1+\tau\left(1-\tau V\right)\left(\uu_{x}^{2}-\uu_{t}^{2}\right)}\;.
\label{eq:squareroot}
\eeq
In order to find a solution $\phi(x,t)$ to (\ref{eq:EoMs}), we proceed by parametrizing it using three generic functions $F$, $X$ and $T$ as follows
\beq
F\left(\uu\right)=X\left(x\right)+T\left(t\right)\;.
\eeq
Then all the derivatives of $\uu$ can be expressed in terms of $F$, $X$ and $T$
\beq
\uu_{x}=\frac{X_{x}}{F'}\;,\,\, \uu_{t}=\frac{T_{t}}{F'}\;,\,\, \uu_{xx}=\frac{X_{xx}}{F'}-X_{x}^{2}\frac{F''}{F'^{3}}\;,\,\, \uu_{tt}=\frac{T_{tt}}{F'}-T_{t}^{2}\frac{F''}{F'^{3}}\;,\,\, \uu_{xt}=-X_{x}T_{t}\frac{F''}{F'^{3}}\;,
\eeq
so that the (\ref{eq:EoMs}) becomes
\begin{eqnarray}
\left(1-\tau V\right)^{2}F'^{2}\left(X_{xx}-T_{tt}\right) - \tau\left(1-\tau V\right)^{3}\left(X_{xx}T_{t}^{2}+T_{tt}X_{x}^{2}\right) \notag \\
= \left(1-\tau V\right)^{2}F''\left(X_{x}^{2}-T_{t}^{2}\right)+\frac{1}{2}\tau V'\left(1-\tau V\right)\left(3+2S\right)F'\left(X_{x}^{2}-T_{t}^{2}\right)+\left(1+S\right)V'F'^{3}\;,
\label{eq:eom1}
\end{eqnarray}
and (\ref{eq:squareroot}) reads
\beq
S^{2}=1+\tau\frac{1-\tau V}{F'^{2}}\left(X_{x}^{2}-T_{t}^{2}\right)\;.
\label{eq:last}
\eeq
We can now solve (\ref{eq:last}) for the combination $X_{x}^{2}-T_{t}^{2}$
and compute its higher order derivatives by chain rule,\footnote{This part relies fundamentally on the fact that the variables are
separate.
} thus obtaining
\beq
X_{x}^{2}-T_{t}^{2}=\frac{S^{2}-1}{\tau\left(1-\tau V\right)}F'^{2}\;,
\label{eq:eom1_2}
\eeq
\beq
X_{xx}=-T_{tt}= \frac{F'\left[2SS'\left(1-\tau V\right)+\tau\left(S^{2}-1\right)V'\right]+2F''\left(S^{2}-1\right)\left(1-\tau V\right)}{2\tau\left(1-\tau V\right)^{2}}F'\;.
\label{eq:constant1}
\eeq
Equation (\ref{eq:constant1}) implies $X_{xx}=-T_{tt}=c_{0}$, where  $c_{0}$ is an arbitrary constant. Setting $c_{0}=0$ and using  (\ref{eq:eom1_2}), equations (\ref{eq:eom1}) and 
(\ref{eq:constant1}) become respectively
\begin{eqnarray}
2 \left(S^2-1\right) \left(1-\tau V\right) F'' + \tau V' F' \left(S+1\right)^2 \left(2S-1\right) &=&0,\;\;\;\;  \label{eq:rr1}\\
2\left(S^{2}-1\right)\left(1-\tau V\right)F''+\left[2SS'\left(1-\tau V\right)+\tau\left(S^{2}-1\right)V'\right]F'&=& 0,\;\;\;\; \label{eq:rr2}
\end{eqnarray}
which can be combined to give
\beq
S'\left(1-\tau V\right)=\tau S\left(S+1\right)V'  \longrightarrow S\left(\uu\right)=\frac{1-c}{c-\tau V\left(\uu\right)} \;,
\label{eq:expS}
\eeq
where $c$ is an arbitrary integration constant.
Plugging expression (\ref{eq:expS}) for $S(\uu)$ into (\ref{eq:rr1}), or equivalently (\ref{eq:rr2}), we obtain the following equation
\beq
2\left(c-\tau V\right)\left(2c-1-\tau V\right)F''+\tau\left(3c-2-\tau V\right)V'F'=0\;,
\eeq
which solution is
\beqa
F'(\uu)&=&\tilde{k}\frac{c-\tau V\left(\uu\right)}{\sqrt{1-2c+\tau V\left(\uu\right)}} \label{eq:Fder}\;,\\ F(\uu)&=&2k\pm\tilde{k}\frac{\left(1+4\tau\kappa\right){\bf F}\left(\frac{    \uu}{2}\vert-\frac{1}{\kappa}\right)-8\tau\kappa {\bf E}\left(\frac{\uu}{2}\vert-\frac{1}{\kappa}\right)}{2\sqrt{\tau \kappa}}.
\label{eq:FF}
\eeqa
In (\ref{eq:FF}), $k$ and $\tilde{k}$ are integration constants and $\kappa$ is related to $c$ via $c=\frac{1}{2}-2\tau\kappa$, while ${\bf F}$ and ${\bf E}$ are
elliptic integrals of the first and second kind, respectively.\\
From the choice $c_0 = 0$ it follows that $X_x = 2\alpha$ and $T_t = 2\beta$ with $\alpha$ and $\beta$ arbitrary constants. Plugging this expression for $X_x$ and $T_t$ together with (\ref{eq:Fder}) into (\ref{eq:last}) one gets the following equation
\beq
\left(\frac{1-c}{c-\tau V\left(\uu\right)}\right)^{2}=1+4\tau\left(1-\tau V (\uu)\right)\left(\alpha^{2}-\beta^{2}\right)\frac{1-2c+\tau V\left(\uu\right)}{\tilde{k}^2\left(c-\tau V\left(\uu\right)\right)^{2}}\;.
\eeq
which allows to fix $\tilde{k}$ as
\beq
\tilde{k}=\pm 2\sqrt{\tau}\sqrt{\alpha^{2}-\beta^{2}}\;.
\eeq
In conclusion,  we have found a class of moving soliton solutions
\beq
\frac{\left(1+4\tau\kappa\right) {\bf F}\left(\frac{\uu}{2}\vert-\frac{1}{\kappa}\right)-8\tau\kappa {\bf E}\left(\frac{\uu}{2}\vert-\frac{1}{\kappa}\right)}{\sqrt{\kappa}}=\pm 2\,\frac{\alpha x+\beta t-k}{\sqrt{\alpha^{2}-\beta^{2}}}\;,
\label{eq:elsolu}
\eeq
which correspond to the $\textsc{T}\bar{\textsc{T}}$ deformation of a particular family of elliptic solutions to the sine-Gordon equation \cite{Hirota1,Faddeev:1974em}. The deformed single kink, is probably the most physically interesting solution belonging to (\ref{eq:elsolu}). With an appropriate scaling of the parameters, we find: 
\beq
8\tau\cos\left(\frac{\uu}{2}\right)+\log\left(\tan\left(\frac{\uu}{4}\right)\right)=\pm2\,\frac{\alpha x+\beta t-k}{\sqrt{\alpha^{2}-\beta^{2}}}\;.
\label{eq:kinksol}
\eeq
\begin{figure}
 \centering
  \subfloat[]{\includegraphics[scale=0.23]{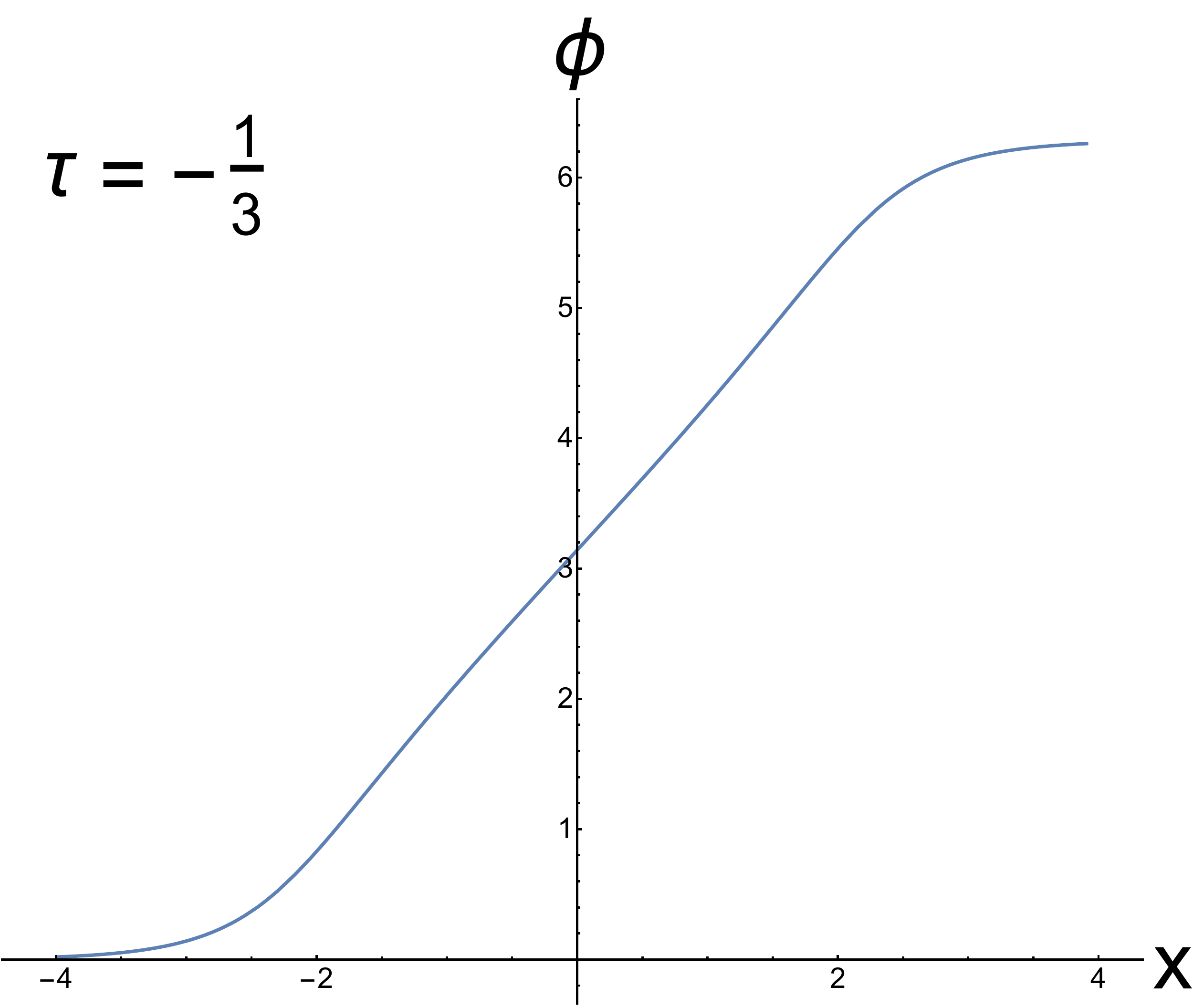}\label{fig:t005}}
  \hspace{1cm}
  \subfloat[]{\includegraphics[scale=0.23]{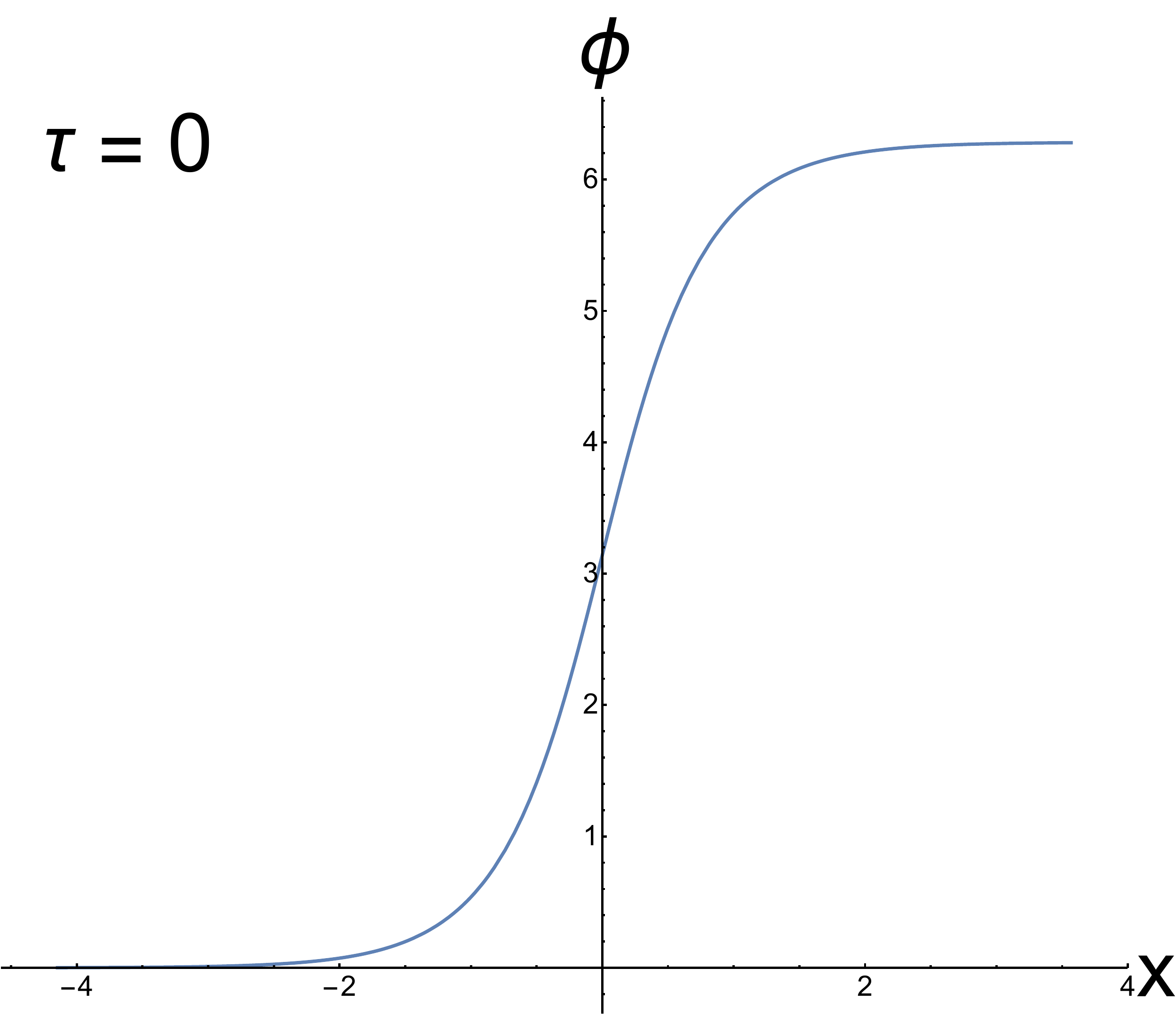}\label{fig:t0}}
  \vspace{0cm}
  \subfloat[]{\includegraphics[scale=0.23]{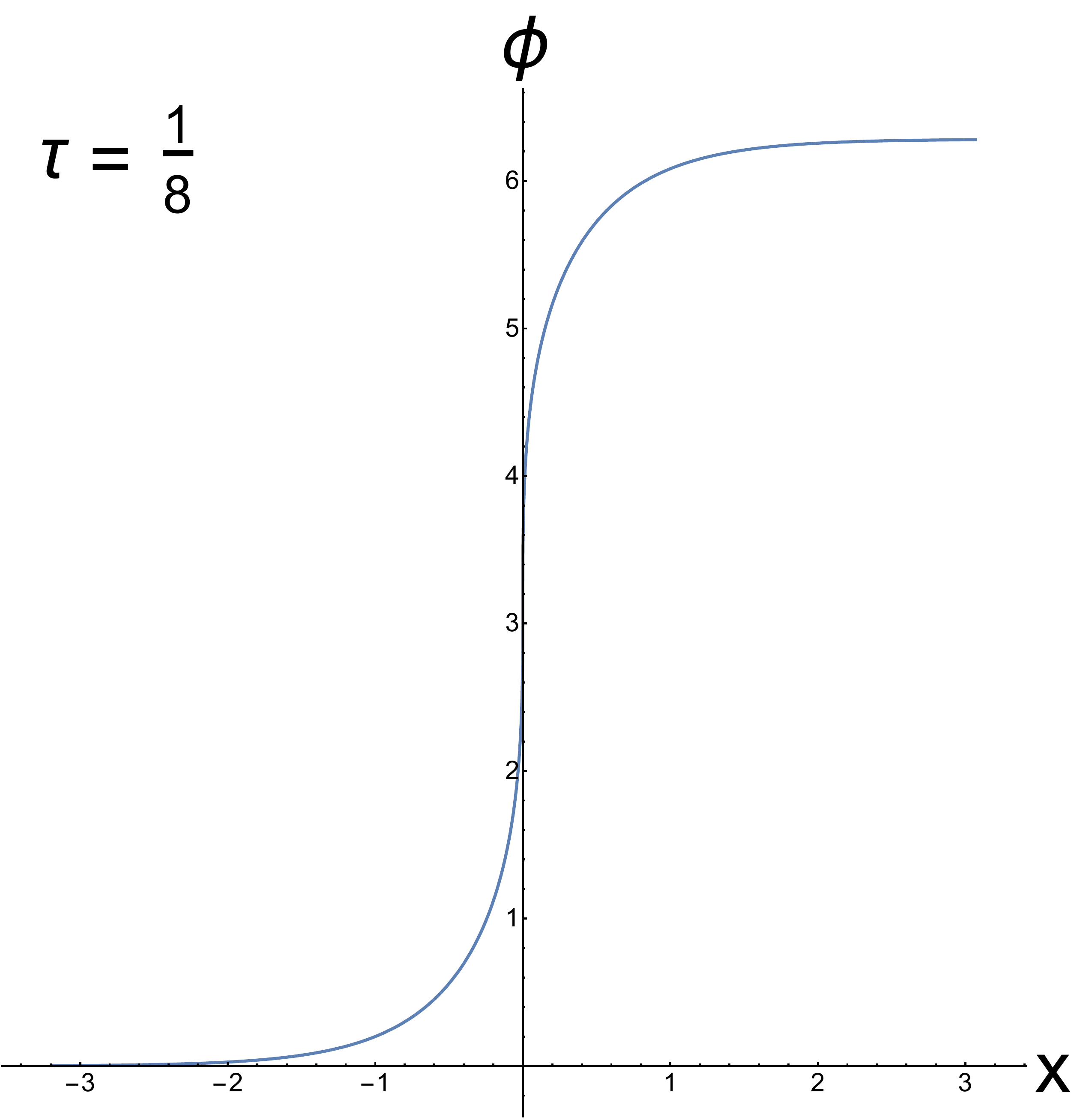}\label{fig:t0125}}
  \hspace{1cm}
  \subfloat[]{\includegraphics[scale=0.23]{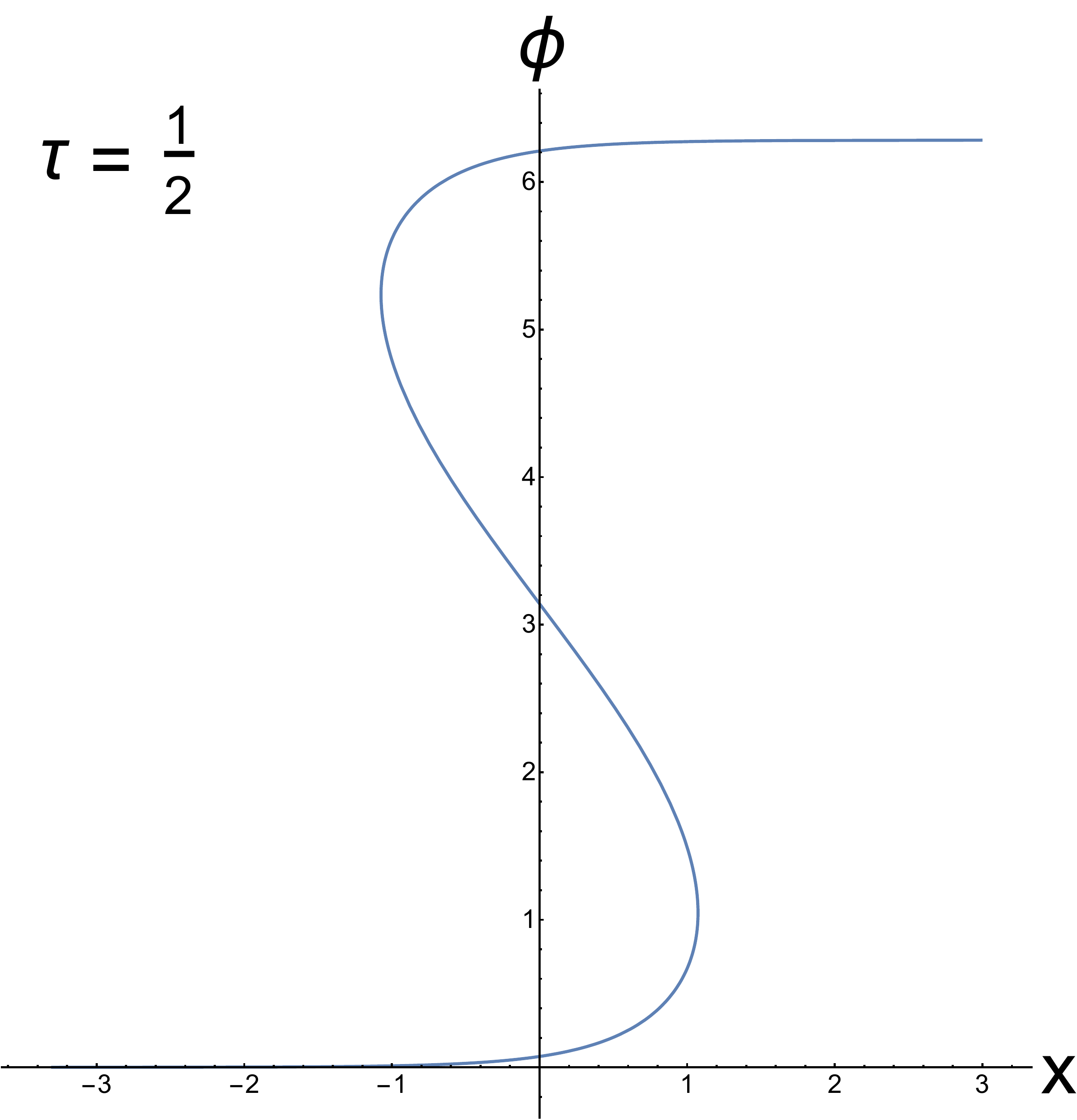}\label{fig:t05}}
  \vspace{0.1cm}
   \caption{The $\textsc{T} \bar{\textsc{T}}$-deformed stationary kink solution (\ref{eq:kinksol}) ($\alpha=1,\beta=0$) for different values of the perturbation parameter $\ta$. The critical value $\ta=1/8$ (c) corresponds to a shock wave singularity.}
  \label{fig:kink}
\end{figure}
In Figure \ref{fig:kink}, the stationary kink-solution is depicted for  four  different  values of the perturbing parameter $\ta$, $\ta=1/8$ corresponds to a shock-wave singularity.  Finally, notice that  (\ref{eq:kinksol}) fulfills
\beq
\begin{cases}
{\displaystyle \partial \uu\left(z,\overline{z}\right)=\frac{2\alpha\sin\left(\frac{\uu\left(z,\overline{z}\right)}{2}\right)}{1-4\tau+4\tau\cos\left(\uu\left(z,\overline{z}\right)\right)}},
\\
{\displaystyle \overline{\partial} \uu\left(z,\overline{z}\right)=\frac{\frac{2}{\alpha}\sin\left(\frac{\uu\left(z,\overline{z}\right)}{2}\right)}{1-4\tau+4\tau\cos\left(\uu\left(z,\overline{z}\right)\right)}}.
\end{cases}\;
\label{eq:backlund0}
\eeq
Since the $\textsc{T} \bar{\textsc{T}}$ perturbation does not spoil integrability, it is tempting to identify  (\ref{eq:backlund0})  as the first-step B\"{a}cklund transformation from the vacuum solution. Unfortunately, equations (\ref{eq:backlund0}) do not contain much information about integrability, and the complete form of the  B\"{a}cklund transformation is expected to be  very complicated.  
A first, more concrete,  step  toward  a fully satisfactory  understanding  of the classical integrability of this system  will be taken  in Section \ref{sec:Lax} below,  where the Lax operators are explicitly  constructed.
Finally, let us conclude this Section with a brief discussion on the more complicated  examples within the family of solutions (\ref{eq:elsolu}). Without much loss in generality  we consider only the stationary  ($\beta=0$, $\alpha=1$) cases.
At $\ta= 0$, equation (\ref{eq:elsolu}) reduces to: 
\beq
x\left(\uu\right)=k\pm\frac{{\bf F}\left(\frac{\uu}{2}\vert-\frac{1}{\kappa}\right)}{\sqrt{\kappa}}\;\longrightarrow\; \uu\left(x\right)=\pm2\,\textrm{\bf am}\left(\sqrt{\kappa}\left(x-k\right)\Big\vert-\frac{1}{\kappa}\right)\;,
\eeq
where $\textrm{\bf am}\left(x\Big\vert k\right)$ is the amplitude of
Jacobi elliptic function, they correspond to  staircase type solutions, see  Figure \ref{fig:Stair}. At $\ta\ne 0$ they display a deformed shape similar to that observed  for the  single kink solution,  with a shock-wave singularities at $\ta \simeq1/8$. 
\subsection{Integrability:  the $\textsc{T}\bar{\textsc{T}}$-deformed  Lax pair}
\label{sec:Lax}
As a first step towards the expression of the Lax operators for the $\textsc{T} \bar{\textsc{T}}$-deformed sine-Gordon model, let us look at the Euler-Lagrange equations in complex coordinates:
\beq
    \partial \left(\frac{\partial \mathcal{L}_{\text{SG}}\left(\phi,\tau\right)}{\partial (\partial \phi)} \right) + \bar{\partial} \left(\frac{\partial \mathcal{L}_{\text{SG}}\left(\phi,\tau\right)}{\partial (\bar{\partial}\phi)} \right) = \frac{\partial \mathcal{L}_{\text{SG}}\left(\phi,\tau\right)}{\partial \phi}\;,
\label{eq:ELequation}
\eeq
with the Lagrangian given by
\beq
    \mathcal{L}_{\text{SG}}\left(\phi,\tau\right) = \frac{V\left(\phi\right)}{1-\tau V\left(\phi\right)} + \frac{-1 + S\left(\phi\right)}{2\tau\left(1-\tau V\left(\phi\right)\right)}\;,\; S(\phi)=\sqrt{1+4\tau\left(1-\tau V\right)\partial\phi\,\bar{\partial}\phi}\;.
\eeq
The potential $V\left(\phi\right)$ is defined in (\ref{eq:sGpotential}), and from the explicit expression of $S$ (we omit the explicit dependence on $\phi$ hereafter) we see that
\beq
    \frac{\partial S}{\partial \phi} = -\ta\,\frac{V'}{1-\ta\,V}\frac{S^2-1}{2S}\;,
\eeq
\beq
    \frac{\partial S}{\partial (\partial \phi)} = \frac{4\tau\left(1-\tau V\right)\bar{\partial} \phi}{2S}\;,\qquad \frac{\partial S}{\partial (\bar{\partial} \phi)} = \frac{4\tau\left(1-\tau V\right)\partial \phi}{2S}\;.
\eeq
Equation (\ref{eq:ELequation}) can be immediately recast into the following form
\beq
    \partial \left( \frac{\bar{\partial} \phi}{S} \right) + \bar{\partial}\left( \frac{\partial \phi}{S} \right) = \frac{V'}{4S}\left(\frac{S+1}{1-\tau V}\right)^2\;.
\label{eq:EOMforLax}
\eeq
With this expression for the  equations of motion, we can proceed and search  for a pair of matrices
\beq
    L = \left(\begin{array}{cc}
         -a & b  \\
         c & a
    \end{array}\right)\;,\qquad \bar{L} = \left(\begin{array}{cc}
         \bar{a} & \bar{b}  \\
         \bar{c} & -\bar{a}
    \end{array}\right)\;,
\eeq
such that the zero-curvature condition
\beq
    \partial \bar{L} - \bar{\partial} L = \left[L,\bar{L}\right]\;,
\label{eq:ZCC}
\eeq
is satisfied iff $\phi$ solves (\ref{eq:EOMforLax}). In terms of the Lax pair's components, (\ref{eq:ZCC}) is equivalent to the following three equations
\begin{subequations}
\label{eq:ZCCcomponents}
\begin{align}
    &\partial \bar{a} + \bar{\partial} a = b \bar{c} - c \bar{b}\;, \label{subeq:ZCCcomponent1}\\
    &\bar{\partial} b - \partial \bar{b} = 2 a \bar{b} + 2 \bar{a} b\;, \label{subeq:ZCCcomponent2}\\
    &\partial \bar{c} - \bar{\partial} c = 2 a \bar{c} + 2 \bar{a} c\;. \label{subeq:ZCCcomponent3}
\end{align}
\end{subequations}
We choose (rather arbitrarily) the first (\ref{subeq:ZCCcomponent1}) to correspond exactly to the equation of motion for $\phi$. It is then reasonable to choose
\beq
    a = \gamma \frac{\partial \phi}{2S} \;,\qquad \bar{a} = \gamma \frac{\bar{\partial} \phi}{2S}\;,
\eeq
with $\kappa$ and arbitrary constant to be determined later. The equations (\ref{eq:ZCCcomponents}) become
\begin{subequations}
\label{eq:ZCCcomponentsstep1}
\begin{align}
    & b \bar{c} - c \bar{b} = \gamma \frac{V'}{8S}\left(\frac{S+1}{1-\tau V}\right)^2\;, \label{subeq:ZCCcomponentstep11}\\
    &\bar{\partial} b - \partial \bar{b} = \gamma \frac{\partial \phi}{S} \bar{b} + \gamma \frac{\bar{\partial} \phi}{S} b\;, \label{subeq:ZCCcomponentstep12}\\
    &\partial \bar{c} - \bar{\partial} c = \gamma \frac{\partial \phi}{S} \bar{c} + \gamma \frac{\bar{\partial} \phi}{S} c\;. \label{subeq:ZCCcomponentstep13}
\end{align}
\end{subequations}
Now it comes the most tricky part of our construction: determining the form of the remaining functions $b$, $c$, $\bar{b}$ and $\bar{c}$. We can proceed by making a perturbative expansion in $\tau$, solving the equations and trying to recognize some pattern in the terms. Sparing the reader the boring details, one arrives at the following Ansatz:
\begin{subequations}
\begin{align}
    &b = \left[\mu e^{\mathbbm i \frac{\phi}{2}}B_+\left(V,S\right)+\tilde{\mu} e^{-\mathbbm i\frac{\phi}{2}} \left(\partial \phi\right)^2 B_-\left(V,S\right)\right]\;, \\
    &c = \left[\frac{1}{\tilde{\mu}} e^{-\mathbbm i \frac{\phi}{2}}B_+\left(V,S\right)+\frac{1}{\mu} e^{\mathbbm i\frac{\phi}{2}} \left(\partial \phi\right)^2 B_-\left(V,S\right)\right]\;, \\
    &\bar{b} = \left[\tilde{\mu} e^{-\mathbbm i \frac{\phi}{2}}B_+\left(V,S\right)+\mu e^{\mathbbm i\frac{\phi}{2}} \left(\bar{\partial} \phi\right)^2 B_-\left(V,S\right)\right]\;, \\
    &\bar{c} = \left[\frac{1}{\mu} e^{\mathbbm i \frac{\phi}{2}}B_+\left(V,S\right) + \frac{1}{\tilde{\mu}} e^{-\mathbbm i\frac{\phi}{2}} \left(\bar{\partial} \phi\right)^2 B_-\left(V,S\right)\right]\;, \\
    & \gamma = \frac{\mathbbm i}{2}\;.
\end{align}
\end{subequations}
Here the parameters $\mu$ and $\tilde{\mu}$ are completely arbitrary complex numbers. They can be, in principle, regarded as two independent spectral parameters. However, as we shortly see, there really exists a single independent spectral parameter, up to global $SL\left(2,\mathbb C\right)$ rotation. The expressions above, when inserted into the equations (\ref{eq:ZCCcomponentsstep1}), give
\beq
    B_+ = \frac{\left(S+1\right)^2}{8S\left(1-\tau V\right)}\;,\;
    B_- = \frac{\tau}{2S}\;.
\eeq
We thus arrive to  the following form of the Lax pair for the $\textsc{T} \bar{\textsc{T}}$-deformed sine-Gordon model:
\begin{align}
\label{eq:sGLaxes}
    L = \left(\begin{array}{cc}
         -\mathbbm i\frac{\partial \phi}{4S} & \mu e^{\mathbbm i\frac{\phi}{2}} \frac{\left(S+1\right)^2}{8S\left(1-\tau V\right)} + \tilde{\mu} e^{-\mathbbm i \frac{\phi}{2}} \left(\partial\phi\right)^2\frac{\tau}{2S}  \\
         \frac{1}{\tilde{\mu}} e^{-\mathbbm i\frac{\phi}{2}} \frac{\left(S+1\right)^2}{8S\left(1-\tau V\right)} + \frac{1}{\mu} e^{\mathbbm i \frac{\phi}{2}} \left(\partial\phi\right)^2\frac{\tau}{2S} & \mathbbm i\frac{\partial \phi}{4S} 
    \end{array}
    \right)\;,\; \notag \\
    \bar{L} = \left(\begin{array}{cc}
         \mathbbm i\frac{\bar{\partial} \phi}{4S} & \tilde{\mu} e^{-\mathbbm i\frac{\phi}{2}} \frac{\left(S+1\right)^2}{8S\left(1-\tau V\right)} + \mu e^{\mathbbm i \frac{\phi}{2}} \left(\bar{\partial}\phi\right)^2\frac{\tau}{2S}  \\
         \frac{1}{\mu} e^{\mathbbm i\frac{\phi}{2}} \frac{\left(S+1\right)^2}{8S\left(1-\tau V\right)} + \frac{1}{\tilde{\mu}} e^{-\mathbbm i \frac{\phi}{2}} \left(\bar{\partial}\phi\right)^2\frac{\tau}{2S} & -\mathbbm i\frac{\bar{\partial} \phi}{4S} 
    \end{array}
    \right)\;. 
\end{align}
There is one final manipulation that we wish to perform. As we mentioned above, the presence of two independent spectral parameters $\mu$ and $\tilde{\mu}$ is redundant and we can fix the dependence of the Lax pair on a single parameter $\lambda = \sqrt{\mu /\tilde{\mu}}$ by applying the following global $SL\left(2,\mathbbm C\right)$ rotation:
\begin{equation}
    L\;\longrightarrow\; \tilde{L} = \mathcal S^{-1} L \mathcal S\;,\qquad \bar{L}\;\longrightarrow  \tilde{\bar{L}} = \mathcal S^{-1} \bar{L} \mathcal S\;,
\end{equation}
where
\begin{equation}
    \mathcal S = \left(\begin{array}{c c}
        \sqrt{\tilde{\mu} \lambda} & 0\\
        0 & \frac{1}{\sqrt{\tilde{\mu} \lambda}}
    \end{array}\right) \equiv \left(\begin{array}{c c}
        \left(\tilde{\mu} \mu\right)^{\frac{1}{4}} & 0\\
        0 & \left(\tilde{\mu} \mu\right)^{-\frac{1}{4}}
    \end{array}\right)\;.
\end{equation}
We end up with the following expressions (omitting the tildas on the transformed Lax operators)
\begin{align}
\label{eq:sGLaxes_singleparameter}
    L = \left(\begin{array}{cc}
         -\mathbbm i\frac{\partial \phi}{4S} & \lambda e^{\mathbbm i\frac{\phi}{2}} \frac{\left(S+1\right)^2}{8S\left(1-\tau V\right)} + \frac{1}{\lambda} e^{-\mathbbm i \frac{\phi}{2}} \left(\partial\phi\right)^2\frac{\tau}{2S}  \\
         \lambda e^{-\mathbbm i\frac{\phi}{2}} \frac{\left(S+1\right)^2}{8S\left(1-\tau V\right)} + \frac{1}{\lambda} e^{\mathbbm i \frac{\phi}{2}} \left(\partial\phi\right)^2\frac{\tau}{2S} & \mathbbm i\frac{\partial \phi}{4S} 
    \end{array}
    \right)\;,\; \notag \\
    \bar{L} = \left(\begin{array}{cc}
         \mathbbm i\frac{\bar{\partial} \phi}{4S} & \frac{1}{\lambda} e^{-\mathbbm i\frac{\phi}{2}} \frac{\left(S+1\right)^2}{8S\left(1-\tau V\right)} + \lambda e^{\mathbbm i \frac{\phi}{2}} \left(\bar{\partial}\phi\right)^2\frac{\tau}{2S}  \\
         \frac{1}{\lambda} e^{\mathbbm i\frac{\phi}{2}} \frac{\left(S+1\right)^2}{8S\left(1-\tau V\right)} + \lambda e^{-\mathbbm i \frac{\phi}{2}} \left(\bar{\partial}\phi\right)^2\frac{\tau}{2S} & -\mathbbm i\frac{\bar{\partial} \phi}{4S} 
    \end{array}
    \right)\;. 
\end{align}
Now, by using the following limiting behaviours
\begin{equation}
    S\;\underset{\tau\rightarrow 0}{\longrightarrow}\; 1\;,\qquad B_+\;\underset{\tau\rightarrow 0}{\longrightarrow}\;\frac{1}{2}\;,\qquad B_-\;\underset{\tau\rightarrow 0}{\longrightarrow}\;0\;,
\end{equation}
we easily verify that, in the vanishing perturbation limit $\tau\rightarrow 0$, we recover, as expected, the usual Lax pair for the sine-Gordon model:
\beq
\label{eq:sGLaxeslimit}
    L = \left(\begin{array}{cc}
         -\mathbbm i\frac{\partial \phi}{4} & \frac{\lambda}{2} e^{\mathbbm i\frac{\phi}{2}}  \\
         \frac{\lambda}{2} e^{-\mathbbm i\frac{\phi}{2}} & \mathbbm i\frac{\partial \phi}{4} 
    \end{array}
    \right)\;,\;
    \bar{L} = \left(\begin{array}{cc}
         \mathbbm i\frac{\bar{\partial} \phi}{4} & \frac{1}{2\lambda} e^{-\mathbbm i\frac{\phi}{2}}  \\
         \frac{1}{2\lambda} e^{\mathbbm i\frac{\phi}{2}}& -\mathbbm i\frac{\bar{\partial} \phi}{4} 
    \end{array}
    \right)\;.
\eeq
Therefore, we have proved that the classical integrability of sine-Gordon model survives the $\textsc{T} \bar{\textsc{T}}$ deformation, by displaying the existence of the Lax pair (\ref{eq:sGLaxes_singleparameter}). We wish to conclude this Section by remarking that the knowledge of the Lax pair for the $\textsc{T} \bar{\textsc{T}}$-deformed sine-Gordon model comes with two additional results:
\begin{itemize}
    \item Single boson BI Lax pair, obtained by simply looking at the Euler-Lagrange equations (\ref{eq:ELequation}) with $V=V'=0$:
\begin{align}
\label{eq:NGLaxes}
    L = \left(\begin{array}{cc}
         -\mathbbm i\frac{\partial \phi}{4S} & 0  \\
         0 & \mathbbm i\frac{\partial \phi}{4S} 
    \end{array}
    \right)\;,\; 
    \bar{L} = \left(\begin{array}{cc}
         \mathbbm i\frac{\bar{\partial} \phi}{4S} & 0 \\
         0 & -\mathbbm i\frac{\bar{\partial} \phi}{4S} 
    \end{array}
    \right)\;. 
\end{align}
    \item sinh-Gordon Lax pair, which can be derived from (\ref{eq:sGLaxes}) by simply redefining the field $\varphi = \mathbbm i \phi$
\begin{align}
\label{eq:shGLaxes}
    L = \left(\begin{array}{cc}
         -\frac{\partial \varphi}{4\tilde{S}} & \lambda e^{\frac{\varphi}{2}} \frac{\left(\tilde{S}+1\right)^2}{8\tilde{S}\left(1-\tau \tilde{V}\right)} - \frac{1}{\lambda} e^{-\frac{\varphi}{2}} \left(\partial\varphi\right)^2\frac{\tau}{2\tilde{S}}  \\
         \lambda e^{-\frac{\varphi}{2}} \frac{\left(\tilde{S}+1\right)^2}{8\tilde{S}\left(1-\tau \tilde{V}\right)} - \frac{1}{\lambda} e^{\frac{\varphi}{2}} \left(\partial\varphi\right)^2\frac{\tau}{2\tilde{S}} & \frac{\partial \varphi}{4\tilde{S}} 
    \end{array}
    \right)\;,\; \notag \\
    \bar{L} = \left(\begin{array}{cc}
         \frac{\bar{\partial} \varphi}{4\tilde{S}} & \frac{1}{\lambda} e^{-\frac{\varphi}{2}} \frac{\left(\tilde{S}+1\right)^2}{8\tilde{S}\left(1-\tau \tilde{V}\right)} - \lambda e^{\frac{\varphi}{2}} \left(\bar{\partial}\varphi\right)^2\frac{\tau}{2\tilde{S}}  \\
         \frac{1}{\lambda} e^{\frac{\varphi}{2}} \frac{\left(\tilde{S}+1\right)^2}{8\tilde{S}\left(1-\tau \tilde{V}\right)} - \lambda e^{-\frac{\varphi}{2}} \left(\bar{\partial}\varphi\right)^2\frac{\tau}{2\tilde{S}} & -\frac{\bar{\partial} \varphi}{4\tilde{S}} 
    \end{array}
    \right)\;, 
\end{align}
where we introduced
\begin{equation}
    \tilde V = 2\left(1-\cosh\varphi\right)\;,\; \tilde S = \sqrt{1-4\tau\left(1-\tau\tilde V\right)\partial\varphi\bar{\partial}\varphi}\;.
\end{equation}
\end{itemize}
This proves that both theories, as expected, retain their integrable structure along the $\textsc{T} \bar{\textsc{T}}$ flow.

\section{Maxwell-Born-Infeld electrodynamics in 4D}
\label{sec:MBI}

Two-photon plane wave  scattering  in  4D Maxwell-Born-Infeld (MBI)  electrodynamics was considered by Schr\"odinger and others in pre-QED  times (see, for example,  \cite{Scharnhorst:2017wzh} for a nice historical review on the early period of non-linear electrodynamics theories). Later,   
in \cite{Barbashov:1967zzz, 1966JETP} it was shown that the scattering of two plane waves in  MBI  electrodynamics can be mapped onto a specific solution of the 2D bosonic BI equations of motion, the $N=2$ model in equations (\ref{eq:Nbos}) and (\ref{eq:NGL}). 
In particular,  it is extremely suggestive that the resulting  phase-shift can be nicely interpreted as being the classical analog  of the  $\textsc{T} \bar{\textsc{T}}$-related scattering phase. Compare,  for example, the results of \cite{Barbashov:1967zzz,1966JETP} with the discussion about the classical origin of the time delay in  \cite{dubovsky2012solving} .

Motivated by these observations, in this Section we investigate the 4D MBI theory of electrodynamics and show that interestingly it shares a lot of common aspects with the 2D bosonic BI models studied in Section \ref{sec:DefLag}. In particular we will see that it arises as a deformation of the Maxwell theory induced by the square root of the determinant of the Hilbert stress-energy tensor.\\
Consider the MBI Lagrangian in 4D defined on a generic background metric $g_{\mu\nu}$ as
\beq
\label{eq:BILgen}
\mathcal{L}^{\textrm{\MBI}}_g(\cA,\tau)=\frac{-\sqrt{|\det\left[ g_{\mu\nu}\right]|}+\sqrt{\det\left[g_{\mu\nu}+\sqrt{2\tau}F_{\mu\nu}\right]}}{2\tau} \;,\; \left(\mu,\nu = \left\lbrace 1,2,3,4 \right\rbrace \right) \;,
\eeq
where $F_{\mu\nu}=\partial_{\mu}\cA_{\nu}-\partial_{\nu}\cA_{\mu}$ is the field strength associated to the abelian gauge field $\cA_{\mu}$. In Euclidean spacetime $(g_{\mu\nu} = \eta_{\mu\nu} \equiv \text{diag}(+1,+1,+1,+1))$, (\ref{eq:BILgen}) takes the  form
\beq
\label{eq:BIL}
\mathcal{L}^{\textrm{\MBI}}(\cA,\ta) = \frac{-1 + \sqrt{1-\ta\,\text{Tr}\left[F^2\right]+\frac{\ta^2}{4}\left(\text{Tr}[F\widetilde{F}]\right)^2}}{2\ta} \;,
\eeq
where $\widetilde{F}_{\mu\nu} = \frac{1}{2}\epsilon_{\mu\nu\rho\sigma}F^{\rho\sigma}$ is the Hodge dual field strength. From the expansion of (\ref{eq:BIL}) in powers of $\ta$ around $\ta=0$
\beqa
\label{eq:BILexp}
\mathcal{L}^{\textrm{\MBI}}(\cA,\ta) &\underset{\ta\rightarrow0}{\sim}& -\frac{1}{4}\textrm{Tr}[F^{2}] + \frac{\ta}{16}\left(\textrm{Tr}[F^2]^{2}-4\textrm{Tr}[F^4]\right) + \mathcal{O}(\ta^2) \notag \\ 
&=&\mathcal{L}^{\Max} + \ta \sqrt{\det[T^{\Max}]} + \mathcal{O}(\ta^2)\;,
\eeqa
one recognizes the Maxwell Lagrangian 
\beq
\mathcal{L}^{\Max}(\cA) = \frac{1}{4} F_{\mu\nu}F^{\mu\nu} = -\frac{1}{4}\textrm{Tr}[F^{2}] \;,
\eeq
at the order $\mathcal{O}(\ta^0)$. 
The $\mathcal{O}(\ta)$ contribution in  (\ref{eq:BILexp})  is instead related to the determinant of the Hilbert stress-energy tensor of the Maxwell theory $T^{\Max}$, which can be computed from the Noether theorem adding the Belinfante-Rosenfeld improvement to make it symmetric and gauge invariant, {\it i.e.}
\beq
\left( T^{\Max} \right)^{\mu\nu}\equiv\frac{\partial\mathcal{L}^{\Max}}{\partial\left(\partial_{\mu}\cA_{\rho}\right)}F^{\nu\rho}-\eta^{\mu\nu}\mathcal{L}^{\Max} = F^{\mu\rho}F^{\nu\rho} - \eta^{\mu\nu}\mathcal{L}^{\Max} \;.
\eeq
Formula (\ref{eq:BILexp}) hints that $\mathcal{L}^{\text{\MBI}}$ may arise from a deformation of Maxwell electrodynamics effected by the operator $\mathcal{O}\equiv\sqrt{\det[T^{\text{\MBI}}]}$ according to the flow equation
\beq
\label{eq:BITTbar}
\partial_\ta \mathcal{L}^{\text{\MBI}} = \sqrt{\det[T^{\text{\MBI}}]} \;,
\eeq
where $T^{\text{\MBI}}$ is the Hilbert stress-energy tensor associated to the MBI Lagrangian. Using the general definition
\beq
\label{eq:TBI}
\left(T^{\text{\MBI}}\right)^{\mu\nu} = \frac{-2}{\sqrt{g}}\frac{\delta\mathcal{L}^{\text{\MBI}}_g}{\delta g_{\mu\nu}} \;,\; \sqrt{g}\equiv \sqrt{|\det[g_{\mu\nu}]|} \;,
\eeq
it is possible to show that, in euclidean spacetime $(g_{\mu\nu}=\eta_{\mu\nu})$, the following relation holds
\beq
\mathcal{O}=\frac{-1+\mathcal{S}(\ta)-2\ta\,\mathcal{L}^{\Max}}{2\ta^{2}\mathcal{S}(\ta)} = \partial_\ta \mathcal{L}^{\text{\MBI}} \;,\; \mathcal{S}(\ta)\equiv \sqrt{\det\left[\eta_{\mu\nu}+\sqrt{2\ta}F_{\mu\nu}\right]} \;,
\eeq
thus proving the validity of (\ref{eq:BITTbar}).\\
As noticed in \cite{Baggio:2018gct}, the presence of an internal symmetry (in the current case the $U(1)$ gauge symmetry) makes the definition of the stress-energy tensor ambiguous. As already appears at the perturbative level in (\ref{eq:BILexp}), here the symmetric and gauge invariant Hilbert stress-energy tensor seems to be the natural choice to get the BI Lagrangian as a deformation of the Maxwell electrodynamics. However let us point out that there is no reason to rule out a priori a deformation induced by the Noether stress-energy tensor, which is neither symmetric nor gauge invariant.\\
Driven by the formal analogy between (\ref{eq:BIL}) and the bosonic 2D BI Lagrangian (\ref{eq:NGL}), now we apply the same strategy of Section \ref{sec:DefLag} to put interactions in the theory.\\
Recasting (\ref{eq:BIL}) into a more compact form
\beq
\label{eq:BILcompact}
\mathcal{L}^{\textrm{\MBI}}(\cA,\ta) = \frac{-1 + \sqrt{1+4\ta\,\mathcal{L}^{\Max}(\mathcal{A})+4\ta^2\mathcal{B}^{\MBI}}}{2\ta} \;,\; \mathcal{B}^{\text{\MBI}} = \det[F] \;,
\eeq
one immediately see that the quantity
\beq
\mathcal{L}^{\text{\MBI}}_\x(\cA,\ta) = \frac{1}{\x} \mathcal{L}^{\text{\MBI}}\left(\cA, \frac{\ta}{\x^2} \right) \;,
\eeq
where $\x$ is again an auxiliary adimensional parameter, satisfies the inhomogeneous Burgers equation
\beq
\label{eq:BurgersBIL}
\partial_\ta \mathcal{L}^{\text{\MBI}}_\x(\cA,\ta) = \mathcal{L}^{\MBI}_\x(\cA,\ta)\,\partial_\x \mathcal{L}^{\MBI}_\x(\cA,\ta) + \frac{\mathcal{B}^{\MBI}}{\x^3} \;,
\eeq
with boundary condition 
\beq
\mathcal{L}^{\MBI}_{\x}(\cA,0) = \frac{1}{\x}\mathcal{L}^{\Max}(\cA) \;.
\eeq
Now it is straightforward to introduce interactions in the theory. Starting from a boundary condition of the form
\beq
\mathcal{L}^{{\MBI,V}}_\x(\cA,0) = \frac{1}{\x}\mathcal{L}^{\Max}(\mathcal{A}) + \x\,V(\cA) \;,
\eeq
where $V(\cA)$ is a derivative-independent potential\footnote{For instance $V$ could be a mass term of the form $V(\cA) = m^2 \cA_\mu \cA^\mu$ which gives the Proca Lagrangian describing a massive spin-$1$ field $\cA_\mu$.}, the solution to (\ref{eq:BurgersBIL}) becomes
\beq
\label{eq:BILV}
\mathcal{L}^{{\text{\MBI}},V}_\x(\cA,\ta) =
\frac{\x\,V}{1-\ta\,V} + \frac{\x}{2\bar{\ta}}\left( -1+\sqrt{\det\left[\eta_{\mu\nu}+\sqrt{\frac{2\bar{\ta}}{\x^2}}F_{\mu\nu}\right]} \right) \;,
\eeq
where $\bar{\tau} = \ta( 1-\ta\,V(\cA) )$ is the usual (local)  redefinition of the deformation parameter. A posteriori it is easy to check that $\mathcal{L}^{{\text{\MBI}},V}_{\x=1}(\mathcal{A},\ta)$ is indeed solution to (\ref{eq:BITTbar}), {\it i.e.}
\beq
\sqrt{\det [T^{\text{\MBI},V}]} = 
-\frac{\mathcal{S}(\bar{\ta}) (2 \bar{\ta}\,V-1)-(2 \ta\,V-1) \left(1+2 \bar{\ta}\,   \mathcal{L}^{\Max}\right)}{2 \bar{\ta}^2\,\mathcal{S}(\bar{\ta})} =
   \partial_\ta \mathcal{L}^{\MBI,V}(\mathcal{A},\ta) \;.
\eeq
Following Section \ref{sec:DefLag}, it is interesting to perform a Legendre transformation on $\mathcal{L}^{\MBI, V}(\mathcal{A},\ta)$ to get the Hamiltonian density $\mathcal{H}^{\MBI,V}(\Pi,\mathcal{A},\ta)$. Again, using a shorthand notation for the time derivative $\dot{\mathcal{A}_\mu}=\partial_4 \mathcal{A}_\mu$, the conjugated momentum is
\beq
\Pi^i = \frac{\partial\mathcal{L}^{\MBI,V}(\mathcal{A},\ta)}{\partial\dot{\mathcal{A}}_i} \;,\; \Pi^4\equiv 0 \;,\; (i=1,2,3) \;,
\eeq 
and the Hamiltonian density takes the form
\beq
\mathcal{H}^{\MBI,V}(\Pi,\mathcal{A},\ta) =
\frac{V(\mathcal{A})}{1-\ta\,V(\mathcal{A})} + \frac{1}{2\bar{\tau}} \left( -1 + \sqrt{1 +4\bar{\tau}\,\mathcal{H}^{\Max}(\Pi,\mathcal{A}) + 4\bar{\tau}^2\,|\vec{\mathcal{P}}^{\MBI}(\Pi,\mathcal{A})|^2} \right) \;,
\eeq
where $\mathcal{H}^{\Max}(\Pi,\mathcal{A}) = -\frac{1}{2}\Pi_i\Pi^i + \frac{1}{4}F_{ij}F^{ij} = -T^{\Max}_{44}$ is formally the Hamiltonian density of the Maxwell theory and $\mathcal{P}^{\MBI}_i(\Pi,\mathcal{A}) = -{\mathbbm i}\,T^{\MBI}_{4i} \;,\; (i=1,2,3)$ , is the $i$-th component of the conserved momentum density of the deformed theory, following the same convention of Section \ref{sec:DefLag}. Notice that $\mathcal{H}^{\MBI,V}(\Pi,\mathcal{A},\ta)$ is formally identical to the Hamiltonian density reported in Section (\ref{sec:DefLag}) for the 2D bosonic theory, and again it satisfies an analogous inhomogeneous Burgers equation.\\

Furthermore, let us stress that setting a field-independent constant potential $V(\mathcal{A})= F_0$, also in this case there exists a special value of the parameter $\ta$, {\it i.e.} $\ta_0=\frac{1}{2F_0}$, such that the determinant of the Hilbert stress-energy tensor takes a constant value
\beq
\det[T^{\MBI}(\ta_0)]= \left(\frac{\pi}{2 \bar{\ta}_0} \right)^4 \;,\; \bar{\ta}_0= \ta_0( 1-\ta_0\,F_0) \;.
\eeq
Finally, we would like to make some comments about the generalization of the $\textsc{T} \bar{\textsc{T}}$ deformation to higher dimensions. Here we found that a 4D theory arises as a deformation induced by a power $1/2$ of the determinant of the stress-energy tensor. This result apparently does not agree with the generalization to higher dimensions proposed in \cite{Cardy:2018sdv}, from which one would expect a power $1/(D-1) = 1/3$ instead. Interestingly, notice also that the operator $\sqrt{\det[T^{\MBI}]}$ can be written in this form
\beq
\sqrt{\det[T^{\MBI}]} = \frac{1}{4}\left( \frac{1}{2}\mathrm{Tr}\left[T^{\MBI}\right]^2 - \mathrm{Tr}\left[\left( T^{\MBI}\right)^2\right] \right) \;,
\eeq
which strongly resembles the generalization of the $\textsc{T} \bar{\textsc{T}}$ operator to higher dimensions recently proposed in \cite{Taylor:2018xcy}, except for the factor $1/2$ in front of $\mathrm{Tr}\left[T^{\MBI}\right]^2$ instead of $1/(D-1) = 1/3$. 

Although in this Section we have seen that there are many similarities at the classical level between the 4D Maxwell-Born-Infeld model and the 2D bosonic model discussed in Section \ref{sec:DefLag}, the situation at the quantum level is in principle much more complicated. However it would be remarkable if a structure similar to that reviewed in Section \ref{sec:DefLag} could emerge for the quantized energy spectrum.

\section{Deformed 2D Yang-Mills}
\label{eq:DefYM2}
The 4D electrodynamics case turns out to be quite special, since in other dimensions the MBI Lagrangian seems not to arise from a deformation of the Maxwell theory driven by any power of the determinant of the Hilbert stress-energy tensor. Solving perturbatively equation (\ref{eq:LagTT0}), with initial condition the Maxwell Lagrangian at $\ta=0$, only for the two-dimensional case we were able to recover the full analytic expression for the deformed Lagrangian: 
\beq
\mathcal{L}^{\Max_2}(\cA,\ta) = \frac{3}{4 \ta} \left(\, _3F_2\left(-\frac{1}{2},-\frac{1}{4},\frac{1}{4};\frac{1}{3},\frac{2}{3};\frac{256}{27} \,\ta\,\mathcal{L}^{\Max_2}(\cA,0)\right) -1 \right) \;,
\label{eq:LYM2}
\eeq
where $\mathcal{L}^{\Max_2}(\cA,0) = \frac{1}{2} F_{21} F^{21}$ is the 2D Maxwell Lagrangian, and $F_{21}=-F_{12}$ is the only non-vanishing component of the field strength. Expression (\ref{eq:LYM2}) is unexpectedly complicated, however, since the quantized energy spectrum should still  satisfy the Burgers equation (\ref{eq:Burgers}), simplifications   may appear at the level of the classical Hamiltonian density.
As before, denoting the time derivative as $\dot{\mathcal{A}}_\mu = \partial_2 \cA_\mu$, the conjugated momenta are
\beq
\label{eq:conjmom}
\Pi^1 = \frac{\partial\mathcal{L}^{\Max_2}(\cA,\ta)}{\partial \dot{\mathcal{A}}_1} \;,\; \Pi^2 = 0 \;,
\eeq
and the explicit form of the Legendre map can be obtained using the Lagrange inversion theorem to invert the relation (\ref{eq:conjmom}). One finds that $F_{21}$ can be expressed in terms of $\Pi^1$ as
\beq
F_{21} = \frac{4\Pi^1}{\bigl(2+\ta\,(\Pi^1)^2\bigr)^2} \;,
\eeq
and "surprisingly" the Hamiltonian density takes a very simple form
\beq
\mathcal{H}^{\Max_2}(\Pi,\ta) = \frac{\mathcal{H}^{\Max_2}(\Pi,0)}{1-\ta\,\mathcal{H}^{\Max_2}(\Pi,0)} \;,
\label{eq:HYM2} 
\eeq
where $\mathcal{H}^{\Max_2}(\Pi,0) = -\frac{1}{2}(\Pi^1)^2 = -T^{\Max_2}_{22}$ is the 2D Maxwell Hamiltonian.
The  results (\ref{eq:LYM2}) and (\ref{eq:HYM2}) can be straightforwardly generalized to encompass the non-abelian 2D Yang-Mills (YM$_2$) theory with generic gauge group $G$. 
In fact, using the following definition for the Hilbert stress-energy tensor of the YM theory
\beq
\left( T^{\YM} \right)^{\mu\nu}\equiv\frac{\partial\mathcal{L}^{\YM}}{\partial\left(\partial_{\mu}\cA_{\rho}^a\right)}F_a^{\nu\rho}-\eta^{\mu\nu}\mathcal{L}^{\YM} \;,
\eeq
where $\mathcal{L}^{\YM}(\cA^a) = \frac{1}{4}F_{\mu\nu}^a F_a^{\mu\nu}$ is the YM Lagrangian and $F_{\mu\nu}^a = \partial_\mu \cA_\nu^a - \partial_\nu \cA_\mu^a + f^{abc}\cA_\mu^b \cA_\nu^c$ is the field strength associated to the non-abelian gauge field $\cA_\mu^a$, it is easy to prove that the deformed non-abelian Lagrangian and Hamiltonian densities, {\it i.e.} $\mathcal{L}^{\YM_2}(\cA^a,\ta)$ and $\mathcal{H}^{\YM_2}(\Pi^a,\ta)$ , have again  the form (\ref{eq:LYM2}) and (\ref{eq:HYM2}) respectively with the formal replacement:
\beq
\mathcal{L}^{\Max_2}(\cA) \rightarrow \mathcal{L}^{\YM_2}(\cA^a) \;,\; \\ \mathcal{H}^{\Max_2}(\Pi) \rightarrow \mathcal{H}^{\YM_2}(\Pi^a) \;,
\eeq
where $\mathcal{L}^{\YM_2}(\cA^a) = \frac{1}{2} F_{21}^a F^{21}_a$ and $\mathcal{H}^{\YM_2}(\Pi^a) = -\frac{1}{2} \Pi^{1\,a} \Pi^1_a = -T^{\YM_2}_{22}$ are the Lagrangian and Hamiltonian density of YM$_2$ respectively.
Although the deformed Lagrangian is very complicated, the Hamiltonian $\mathcal{H}^{\YM_2}(\Pi^a,\ta)$ fulfills
\beq
\partial_\ta \mathcal{H}_\x^{\YM_2}(\Pi^a,\ta) = \mathcal{H}_\x^{\YM_2}(\Pi^a,\ta)\,\partial_\x \mathcal{H}_\x^{\YM_2}(\Pi^a,\ta) \;, 
\label{eq:BHYM}
\eeq
with initial condition $\mathcal{H}_\x^{\YM_2}(\Pi^a,0) = \x\,\mathcal{H}^{\YM_2}(\Pi^a)$, which means that $\mathcal{H}^{\YM_2}(\Pi^a,\ta)$ behaves, under the  $\textsc{T} \bar{\textsc{T}}$  deformation, as a pure potential term (cf. Section \ref{sec:DefLag}).  The latter  property can be interpreted as an explicit manifestation  of  the well known  pure topological character of YM$_2$. 

This simple observation  directly motivated the following  proposal for the deformed versions of the partition functions/heat kernels  \cite{Migdal:1975zg,Rusakov:1990rs,Caselle:1993mq,Gross:1994ub} which is compatible with all known consistency constraints \cite{Smirnov:2016lqw, Cavaglia:2016oda, Cardy:2018sdv}.  The partition function of  YM$_2$ defined on an orientable 2D manifold $\cM$ with genus $p$ and metric $g_{\mu \nu}$  is
\beq
Z^{\cM}(A) = \int {\cal D} {\cal A}_\mu \; e^{-\frac{1}{4 \tilde{g}^2} \int_{\cM}  dx^2 \sqrt{g}\, {\text  Tr}[F_{\mu \nu}^a F^{\mu \nu}_a]} = \sum_{\cR} d_{\cR}^{2 - 2p} e^{- \frac{\tilde{g}^2}{2}  A\,\bC_2(\cR) }\;,
\label{eq:ZYM2}
\eeq
where we have restored the explicit dependence on the  Yang-Mills coupling constant $\tilde g$.
In (\ref{eq:ZYM2}),  $A$ is the total area of $\mathcal{M}$, the sum is over all equivalence classes of irreducible representations ${\cal R}$ of the gauge group
$G$, $d_\cR$  is their dimension and $\bC_2(\cR)$ is the quadratic Casimir in the representation $\cR$. The generalization of (\ref{eq:ZYM2}) to a manifold with genus $p$ and $n$ boundaries corresponds to the so-called heat kernel:
\beq
Z^{\cM}(g_1,\dots, g_n |A) =  \sum_{\cR} d_{\cR}^{2 - 2p -n } \chi_\cR (g_1) \dots \chi_\cR(g_n) e^{- \frac{\tilde{g}^2}{2}  A\,\bC_2(\cR) }\;,
\label{eq:HKYM2}
\eeq
where $g_i$ are the Wilson loops evaluated along the boundaries, and $\chi_\cR$ denotes the Weyl character of the representation $\cR$. According to  (\ref{eq:HYM2}), the  $\textsc{T} \bar{\textsc{T}}$ contribution is then  included through a simple redefinition, in the heat kernel (\ref{eq:HKYM2}), of the  eigenvalues of the  quadratic Casimir operator:
\beq
\bC_2(\cR) \rightarrow \bC_2({\cal R},\ta)= \frac{\bC_2({\cal R})}{1-\ta \,\frac{\tilde{g}^2}{2}\, \bC_2({\cal R}) } \;,
\label{eq:Casimirtau}
\eeq
where the dressed operator $\bC_2({\cal R},\ta)$, also fulfills equation (\ref{eq:BHYM}). Since (\ref{eq:HKYM2}) depends only on the surface area $A$ of the manifold,  the deformed version $Z^{\cM}(g_1,\dots, g_n;\ta |A)$ satisfies 
\beq
-\partial_\ta Z^{\cM}(g_1,\dots, g_n;\ta |A) = A\,\partial_A^2 Z^{\cM}(g_1,\dots, g_n;\ta |A) \;.
\eeq
With  the prescription (\ref{eq:Casimirtau}), all the diffusion-type  relations introduced in \cite{Cardy:2018sdv} (see also \cite{Dubovsky:2018bmo, Datta:2018thy}) for the partition functions on various geometries are automatically fulfilled:
\begin{itemize}
\item  {\bf Cylinder}: The  cylinder partition function $Z^{\text{Cyl}}(g_1, g_2 |A)$  corresponds to the $n=2$, $p=0$ case of (\ref{eq:HKYM2}). 
Setting $A =  R L$, and implementing the prescription (\ref{eq:Casimirtau}), $Z^{\text{Cyl}}(g_1, g_2;\ta |A)$ trivially satisfies Cardy's equation:
\beq
-\partial_\ta Z^{\text{Cyl}}(g_1, g_2;\ta |A) = (\partial_L -1/L) \partial_R  Z^{\text{Cyl}}(g_1, g_2;\ta |A) \;.
\eeq
\item{\bf Torus}:
The partition function on the torus, $Z^{\text{T}}(A)$  corresponds to the $n=0$, $p=1$ case of (\ref{eq:HKYM2}) with  $A=L_1 L_2'-L_2 L_1'$, while the consistency equation for the deformed   partition function is:
\beq
-\partial_\ta Z^{\text{T}}(\ta |A) = \left[ \partial_{L_1} \partial_{L_2'} -\partial_{L_2}  \partial_{L_1'} - \frac{1}{A} \left(  L_1 \partial_{L_1} +  L'_1 \partial_{L'_1} + L_2 \partial_{L_2} +  L'_2 \partial_{L'_2} \right) \right] Z^{\text{T}}(\ta |A) \;.
\eeq
\item{\bf Disk and Cone}:
In the case of a disk, or more in general of a cone with opening angle $\mathcal{X}$, the deformed partition function $Z^{\text{Cone}}(g_1;\ta |A)$ corresponding to $n=1$, $p=0$ and area $A=\frac{1}{2}\,\mathcal{X}R^2$ satisfies
\beq
-\partial_\ta Z^{\text{Cone}}(g_1;\ta |A) = \frac{1}{R}\,\mathcal{X}\partial_\mathcal{X}\left( \frac{1}{\mathcal{X}}\partial_R Z^{\text{Cone}}(g_1;\ta |A) \right) \;.
\eeq
\end{itemize}
Finally, let us stress again that the modification (\ref{eq:Casimirtau}) in (\ref{eq:HKYM2}) is expected to hold in general for any value of $p$ and $n$, possibly leading to a consistent deformation of the whole YM$_2$ setup.
\section{Conclusions}
\label{sec:conclusions}

The Maxwell-Born-Infeld  model is  still  playing an important role in modern theoretical  physics. It was initially proposed as a generalization of electrodynamics, in the attempt to impose an upper limit on the electric field of a point charge, and it corresponds to  the only non-linear extension of Maxwell equations that ensures the absence of birefringence and shock waves. Another   important feature of this  special non-linear field theory is its electric-magnetic self-duality.

The Maxwell-Born-Infeld theory  emerges, from this work,  as a natural 4D generalization of the $\textsc{T} \bar{\textsc{T}}$-deformed  2D models, as  it shares with them some of the properties that make  this perturbation so interesting.
There are many aspects that deserve further investigation. First of all, it would be nice to extend the ideas of  \cite{Cardy:2018sdv} to this 4D theory and try to derive an evolution-type  equation for the quantum energy spectrum at finite volume. 

It would be important to explore the classical and quantum properties of the models corresponding to the deformed Lagrangians  (\ref{eq:BILV}) and  to extend the analysis to  more general gauge theories.

Considering the interpretation of the 2D examples within the $AdS_3/$CFT$_2$ framework given in \cite{McGough:2016lol}, the search for  analog  deformations
that preserve integrability  in the ABJM model and $\mathcal{N} = 4$ super Yang-Mills, could lead to  important progresses in  our understanding  of quantum gravity.

Investigating, at a deeper level, the  geometrical meaning of the $\textsc{T} \bar{\textsc{T}}$ deformation in the 2D setup by continuing the study of classical  integrable models started in Section \ref{sec:sine} appears to be a more feasible but  equally important objective. We have now a good control on the deformed quantum spectrum but we have not yet reached an equally satisfactory level of understanding  about the influence that this deformation has on classical solutions such as multi-kink or breather configurations.  Adapting   B\"acklund's, Hirota's and the Inverse Scattering methods to the current  setup would correspond to a natural extension of some of the results presented in this paper. Finally, it is  important to proceed with  some concrete application of the YM$_2$  heat kernel proposal of Section \ref{eq:DefYM2} and in particular with the study  of the large N limit, which  might display novel physical and mathematical features compared to the unperturbed cases.

\section*{Acknowledgements}

We are especially grateful to  Ferdinando Gliozzi and  Sasha Zamolodchikov  for inspiring discussions, Andrea Cavagli\`a for help at the early stages of this project and for useful comments. 
This project was partially supported by the INFN project SFT, the EU network GATIS+,
NSF Award  PHY-1620628, and by the FCT Project PTDC/MAT-PUR/30234/2017 "Irregular connections on algebraic curves and Quantum Field Theory

\bibliography{Biblio3}

\end{document}